\documentclass[%
reprint,
superscriptaddress,
amsmath,amssymb,
aps,
prl,
]{revtex4-2}
\usepackage{makeidx}
\usepackage[unicode=true, colorlinks=true, linkcolor=blue, citecolor=blue, urlcolor=blue]{hyperref}
\makeindex
\usepackage{graphicx}% Include figure files
\usepackage{dcolumn}% Align table columns on decimal point
\usepackage{bm}% bold math
\usepackage{mathtools}
\usepackage{siunitx}  
\usepackage{amsmath}  
\usepackage{booktabs} 
\usepackage{array}    
\usepackage{multirow} 
\usepackage{makecell}
\usepackage{mathtools}

\usepackage{xr-hyper}         
\usepackage{hyperref}     
\begin{document}

\preprint{APS/123-QED}

\title{High-Performance Fully Passive Discrete-State Continuous-Variable Quantum Key Distribution With Local Local Oscillator}

\author{Yu Zhang}
\affiliation{State Key Laboratory of Quantum Optics Technologies and Devices, Institute of Opto-Electronics, Shanxi University, Taiyuan 030006, China}
\author{Xuyang Wang}
\email{wangxuyang@sxu.edu.cn}
\affiliation{State Key Laboratory of Quantum Optics Technologies and Devices, Institute of Opto-Electronics, Shanxi University, Taiyuan 030006, China}
\affiliation{Collaborative Innovation Center of Extreme Optics, Shanxi University, Taiyuan 030006, China}
\affiliation{Hefei National Laboratory, Hefei 230088, China}
\author{Chenyang Li}
\email{chenyangli@astri.org}
\affiliation{Hong Kong Applied Science and Technology Research Institute}
\author{Jie Yun}
\affiliation{State Key Laboratory of Quantum Optics Technologies and Devices, Institute of Opto-Electronics, Shanxi University, Taiyuan 030006, China}
\author{Qiang Zeng}
\affiliation{Beijing Academy of Quantum Information Sciences, Beijing 100193, China}
\author{Zhiliang Yuan}
\affiliation{Beijing Academy of Quantum Information Sciences, Beijing 100193, China}
\author{Zhenguo Lu}
\affiliation{State Key Laboratory of Quantum Optics Technologies and Devices, Institute of Opto-Electronics, Shanxi University, Taiyuan 030006, China}
\affiliation{Collaborative Innovation Center of Extreme Optics, Shanxi University, Taiyuan 030006, China}
\author{Yongmin Li}
\email{yongmin@sxu.edu.cn}
\affiliation{State Key Laboratory of Quantum Optics Technologies and Devices, Institute of Opto-Electronics, Shanxi University, Taiyuan 030006, China}
\affiliation{Collaborative Innovation Center of Extreme Optics, Shanxi University, Taiyuan 030006, China}
\affiliation{Hefei National Laboratory, Hefei 230088, China}

\date{\today}% It is always \today, today,

\begin{abstract}
We propose and demonstrate a fully passive discrete-state continuous-variable quantum key distribution (CV-QKD), which can eliminate all modulator side channels on the source side, using a local local oscillator (LLO). The CV-QKD system achieves a maximum transmission length of $100$ km with a repetition rate of $1$ GHz using specially designed phase rotation and discretization methods, and the corresponding secret key bit rate is $127$ kbps, as estimated based on the amplitude of prepared states at the transmitter, as well as the first- and second-order moments of quadratures at the receiver by employing the convex optimization without imposing any assumptions on the quantum channel. The performance of the protocol is similar to that of modulated CV LLO protocols and better than those of passive discrete-variable and CV protocols. Our protocol is expected to play an important role in the quantum metropolitan area networks and quantum access networks with high realistic security.  
\end{abstract}

\maketitle

\textit{Introduction}---Based on the fundamental laws of quantum mechanics, quantum key distribution (QKD) allows two distant parties to establish information-theoretically secure keys in the presence of eavesdroppers. The development of QKD has remarkably progressed since the BB84 protocol was proposed~\cite{RevModPhys.92.025002, Zhang2024, Pirandola:20,RevModPhys.84.621,RevModPhys.81.1301}. For discrete-variable (DV) QKD, the transmission length has been extended to $1200$ km in free space~\cite{Liao2017}, and to $1002$ km in optical fibers~\cite{PhysRevLett.130.210801}. For continuous-variable (CV) QKD, the transmission length has been extended to $202.81$ km with transmitted local oscillator (TLO)~\cite{PhysRevLett.125.010502}, and to $100$ km when using a local local oscillator (LLO)~\cite{Hajomer2024}. CV-QKD protocols are compatible with classical coherent optical communication, and the whole system can be integrated on a photonics chip~\cite{Zhang2019,Li:18,Jia_2023,Bian_2024,Pietri:24,Hajomer:24}. They are promising to perform a crucial role in the quantum metropolitan area networks and quantum access networks~\cite{PhysRevApplied.16.064051,Wang:23,Li:24,Hajomer:2024}.

When QKD protocols are implemented on realistic devices, side channels and device imperfections can compromise their security. Device-independent (DI) QKD protocol that guaranteed by the violation of the Bell inequality can promise the highest level of practical security~\cite{Zhang:22,Wooltorton:24,Tan:24}. However, the stringent requirements and low performance hinder its wide application. A more practical protocol, measurement-device-independent (MDI) QKD, has been proposed and demonstrated to eliminate all side channels in the measurement devices~\cite{Lo:12,Braunstein:12,Liu:13,Yin:16,Cao:20,Li:23,Tian:22,Pirandola:15}. Most state-of-the-art DV- and CV-QKD implementations assume trustworthy sources and perfectly prepared states~\cite{Liao2017,PhysRevLett.130.210801,PhysRevLett.125.010502,Hajomer2024}. However, realistic modulators can admit side channels that might directly leak information to an eavesdropper (Eve) and suffer from Trojan horse attacks~\cite{Gisin:06,Khan:14}. In addition, perfect modulation in state preparation is prevented by many factors, such as the resolution of modulation~\cite{Jouguet:12}, the stability of modulators~\cite{Liu:17}, the correlations between adjacent pulses~\cite{Yoshino:18}, the laser intensity fluctuation~\cite{Laudenbach:18}, and so on. Many effective methods have been proposed to solve the imperfections and quantify the potentially leaked information~\cite{Li:21,Huang:16,Derkach:17,Usenko:10,Derkach:16,Tamaki:16,Pereira:19,Hajomer:22}, but none can simultaneously handle all modulator side channels. A fully passive DV-QKD protocol has been proposed to eliminate modulator side channels at the source~\cite{Wang:2023,Lu:23,Hu:23,Zapatero:23}. However, its key rate is at least an order of magnitude lower than that of its active counterpart. Thermal source based passive CV-QKD can achieve high speed preparation of Gaussian states without active modulation~\cite{Li:24,Qi:20,Qi:18,Huang:21,Wu:21}. It requires high power thermal source and mode overlap to suppress the excess noise of the passive state preparation, furthermore, the photon leakage noises due to the TLO scheme significantly limit the transmission lengths. To address this, passive discrete-state CV-QKD has been developed to achieve comparable secret key rates to active CV-QKD while eliminating modulator-related side channels~\cite{Li2022Passive}.

In this letter, we propose and demonstrate a fully passive discrete-state LLO CV-QKD protocol. The transmitter (Alice) and receiver (Bob) obtain their data via the heterodyne detection of two phase-randomized pulsed fields. Their data are correlated due to the correlated phases of the two pulsed fields that emitting from one laser. By using phase rotation and discretization methods, equivalent discrete distribution of coherent states is achieved. As the protocol requires no active modulators and quantum random number generator, it greatly simplifies the CV-QKD system while showing performance comparable to those of active CV-QKD protocols~\cite{Hajomer2024,Tian:23,Wang:22,Pan:22}, and outperforming reported DV and CV passive protocols. Notably, the secret key rate is directly calculated based on the amplitude of the prepared states on Alice’s side as well as the first- and second-order moments of quadratures on Bob’s side by employing the convex optimization without any assumptions on the quantum channel.

%\section{PASSIVE-SOURCE-BASED Discrete States CV-QKD PROTOCOL }

\begin{figure}[htbp]
	\centering
	\includegraphics[trim=10mm 5mm 5mm 10mm,clip,width=\columnwidth]{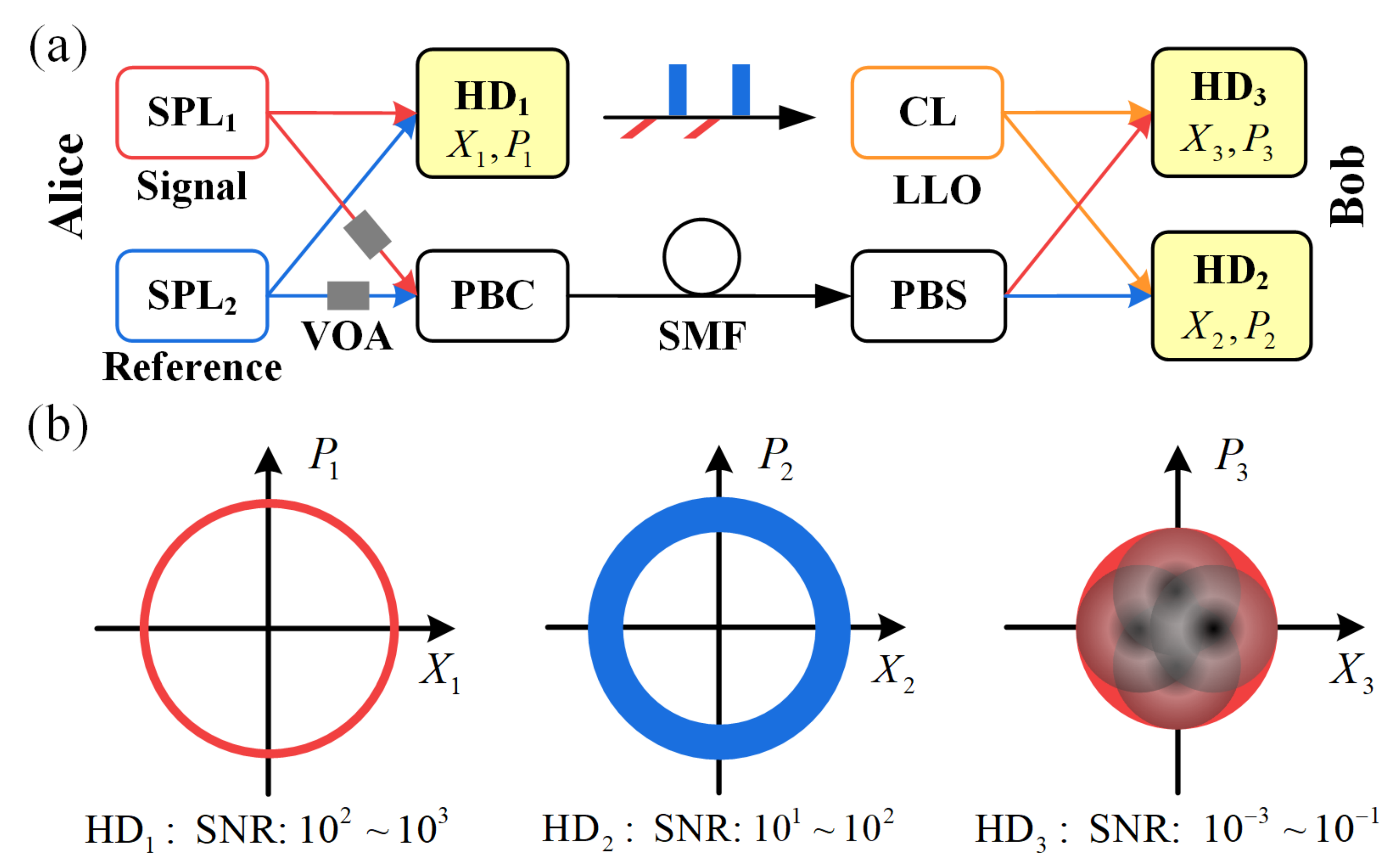}
	\caption{Schematic diagram of the passive discrete-state LLO CV-QKD protocol. SPL: seeded pulse laser; CL: continuous laser; VOA: variable optical attenuator; HD: heterodyne detection; PBC: polarization beam combiner; PBS: polarization beam splitter; SMF: single-mode fiber.}
	\label{fig:1}
\end{figure}

\textit{Passive discrete-state LLO protocol}---In the passive discrete-state LLO protocol as shown in Fig.~\ref{fig:1}(a), the transmitter Alice employs two independent seeded pulse lasers to generate two optical pulse sequences (one noted signal and the other noted reference), whose modes can be overlapped near perfectly~\cite{Comandar:16,Yuan:14,Yuan:2014,Yuan:16}. Parts of signal and reference beams are injected into the heterodyne detector $\text{HD}_1$ and the quadratures of the signal $(X_1,P_1)$ are measured. To minimize the effects of shot noise and electronic noise, strong signal and reference pulses with equal intensity are used. Alice attenuates the intensity of the other part of the signal beam to the single-photon level per pulse (quantum signal), and the other part of the reference beam to an appropriate level according to its influence on quantum signal and the attenuation of the quantum channel. Both attenuated laser pulses are combined at a polarization beam combiner (PBC) and then sent to Bob via a single-mode fiber in polarization and time multiplexing approach.

The receiver Bob demultiplexes the received reference and quantum signal pulses, and directs them into two heterodyne detectors ($\text{HD}_2$ and $\text{HD}_3$), respectively. In the heterodyne detection, both the reference and quantum signal pulses interfere with the continuous LLO beams from one laser, and the measured quadratures are $(X_2,P_2)$ and $(X_3,P_3)$, respectively. Figure~\ref{fig:1}(b) depicts the distribution of three measured states in the phase space. The typical range of signal to noise ratio (SNR) of $\text{HD}_\text{s}$, which are defined as the variance of quadratures versus the variance of noises, are presented. A high SNR in $\text{HD}_1$ can ensure little prepared noise, and a high SNR in $\text{HD}_2$ can ensure the phase of LLO can be processed precisely. Using the phase rotation and discretization methods (detailed in the following section), Bob transforms his phase-randomized states measured in $\text{HD}_3$ into discretely distributed states (black solid circles) according to the quadratures in $\text{HD}_1$ and $\text{HD}_2$. Here, four-state discretization is shown as an example. Other number of discretization is straightforward.

The subsequent parameter estimation process is similar as that of the prepare-and-measure scheme in discrete modulation (DM) protocols~\cite{Lin:19}. Alice publicly announces the amplitude of the prepared states, then Bob estimates the parameters of the experimental system and calculate the secret key rates using Eqs. (\ref{eq:1}) and (\ref{eq:2}). For the DM CV-QKD protocol under linear-channel assumption, the channel transmission $T$ and excess noise $\varepsilon$ are usually estimated to calculate the secret key rate~\cite{Laudenbach:2018}. Herein, the amplitude of the prepared states on Alice's side and the first- and second-order moments of the quadratures measured on Bob's side are directly employed to estimate the secret key rate (Eq.~(\ref{eq:2})). In this case, there are not any assumptions about the quantum channel, and there is also no need to estimate the excess noise on Alice's side.
\begin{figure*}[htbp]
	\includegraphics[trim=5mm 10mm 10mm 5mm,clip,scale=0.45]{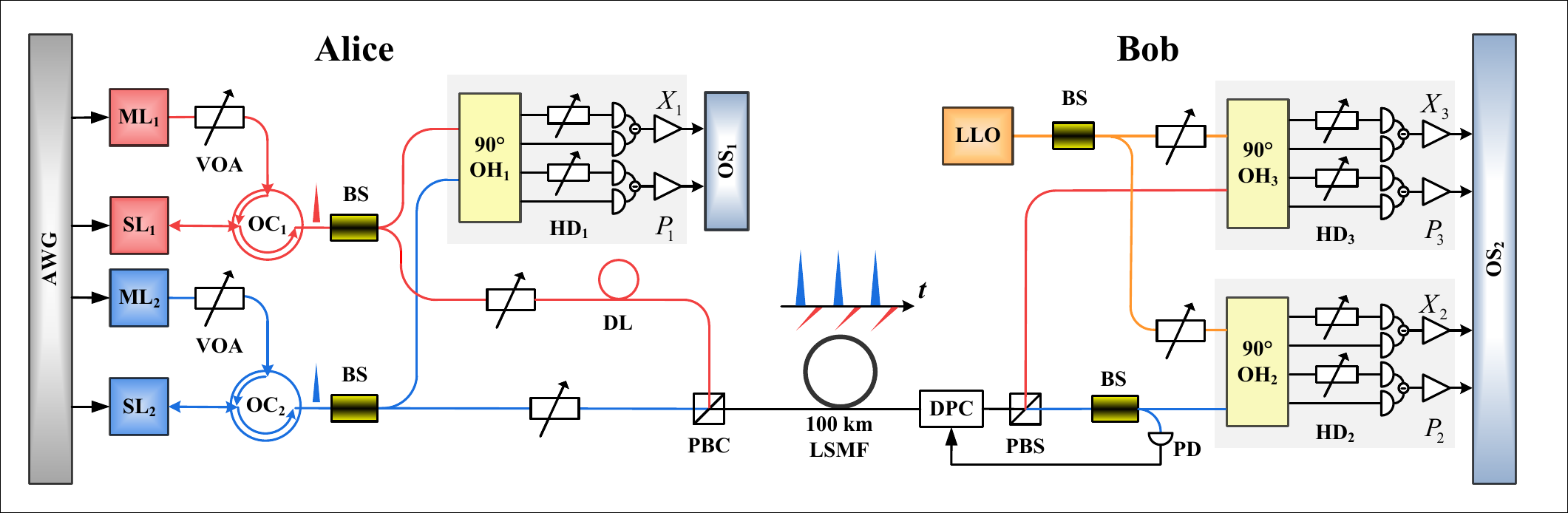}
	\caption{Schematic diagram of the experimental setup. AWG: arbitrary waveform generator; ML: master laser; SL: slave laser; CL: optical circulator; BS: beam splitter; 90° OH: 90° optical hybrid; DL: delay line; DPC: dynamic polarization controller; PD: photodetector; LLO: local local oscillator; 100 km LSMF: 100 kilometer low-loss single-mode fiber; OS: oscilloscope.}
	\label{fig:2}
\end{figure*}

The asymptotic secret key rate in reverse reconciliation is given by~\cite{Lin:20,Winick:18}:
\begin{equation}
	\begin{aligned}
		{{R}^{\infty }}=\underset{{{\rho }_{\operatorname{AB}}}\in \mathcal{S}}{\mathop{\min }}\,\text{ }D\left[ \mathcal{G}\left( {{\rho }_{\operatorname{AB}}} \right)\parallel \mathcal{Z}\left( \mathcal{G}\left( {{\rho }_{AB}} \right) \right) \right]-{{p}_{\text{pass}}}{{\delta }_{\text{EC}}}
	\end{aligned}
	\label{eq:1}
\end{equation}
The set $\mathcal{S}$ contains all density operators ${\rho}_{\operatorname{AB}}$ compatible with experimental observations, where ${\rho }_{\operatorname{AB}}$ presents the joint state of Alice and Bob. The parameter $p_{\text{pass}}$ represents the sifting probability, and is 100$\%$. The term $\delta_{\text{EC}}$ is the information leaked during the error-correction step. The term $D(\rho\parallel\sigma)$ represents the quantum relative entropy, which can be calculated using the convex optimization method with Eq.(\ref{eq:2}), where the first- and second-order moments of the quadratures measured by Bob are used as the constraint condition. 
\begin{equation}
	\begin{aligned}
		&\text{minimize} \quad D(\mathcal{G}(\rho_{AB}) \| \mathcal{Z}[\mathcal{G}(\rho_{AB})]) \\
		&\text{subject to} \\
		&\quad \quad \quad \quad \quad \mathrm{Tr}[\rho_{AB}(|k\rangle\langle k|_A \otimes \hat{X})] = p_k \langle \hat{X}_k \rangle, \\
		&\quad \quad \quad \quad \quad \mathrm{Tr}[\rho_{AB}(|k\rangle\langle k|_A \otimes \hat{P})] = p_k \langle \hat{P}_k \rangle, \\
		&\quad \quad \quad \quad \quad \mathrm{Tr}[\rho_{AB}(|k\rangle\langle k|_A \otimes \hat{X}^2)] = p_k \langle \hat{X}_k^2 \rangle, \\
		&\quad \quad \quad \quad \quad \mathrm{Tr}[\rho_{AB}(|k\rangle\langle k|_A \otimes \hat{P}^2)] = p_k \langle \hat{P}_k^2 \rangle, \\
		&\quad \quad \quad \quad \quad \mathrm{Tr}[\rho_{AB}] = 1, \\
		&\quad \quad \quad \quad \quad \mathrm{Tr}_B[\rho_{AB}] = \sum_{i,j=0}^3 \sqrt{p_i p_j} \langle \varphi_j | \varphi_i \rangle |i\rangle \langle j|_A, \\
		&\quad \quad \quad \quad \quad \rho_{AB} \geq 0. 		
		\label{eq:2}
	\end{aligned}
\end{equation}

The subsequent steps of the protocol follow the standard data processing procedures, including reverse reconciliation and privacy amplification \cite{Jeong:22,Feng:23,Yang:17}.

%\section{EXPERIMENTAL SETUP}  %Sec 3

\textit{Experimental steup}---Figure~\ref{fig:2} illustrates the experimental setup of our passive discrete-state LLO CV-QKD system. Both the seeded pulse laser sources on Alice’s side are composed of a gain-switched master laser (ML), a variable optical attenuator (VOA), an optical circulator (OC), and a gain-switched slave laser (SL). A four-channel arbitrary waveform generator (AWG) is employed to drive the lasers to generate two synchronous pulse sequences with a repetition rate of $1$ GHz. The ML produces phase-randomized pulses that serve as the seeds for the SL to generate the short pulses. The widths of the electrical pulses used to drive the ML and SL are $300$ and $200$ ps, respectively. The intensities of laser pulses injected into and output from the SL are $\sim 10^6$ and $\sim 10^7$ photons per pulse respectively. This method can reduce the emission time jitter and enable frequency chirp synchronization while maintaining random optical phases of the emitted laser pulses. The center wavelength of all lasers was tuned to $1550.12$ nm using temperature controllers. The observed second-order intensity correlation value of interference between the independent signal and reference pulses is $g^{(2)} = 1.47$, almost reaching the theoretical limit of $1.50 $~\cite{Comandar:16}.

The intensities of the signal and reference pulses incident on $\text{HD}_1$ are both $\sim 10^6$ photons per pulse. The output paths of the $90^\circ$ optical hybrid ($90^\circ$ OH) are equipped with two VOAs to balance the two arms of the homodyne detectors. The measurement results were acquired by a high-speed oscilloscope (OS). At the input ports of the PBC, the intensities of quantum signal pulses and reference pulses are $\sim 1$ and $\sim 10^4$--$10^6$ photons per pulse, respectively. 

The performance of the system was demonstrated on four transmission fibers with different lengths ($2$ m, $25$ km, $50$ km, and $100$ km). In the case of $2$ m fiber, the transmitter and receiver were directly connected to calibrate the amplitude of the prepared states and achieve a maximum secret key rate. The $25$ km and $50$ km fibers are standard single-mode fibers and the $100$ km fiber is a low-loss single-mode fiber (0.16 dB/km).

At the receiver side, Bob demultiplexes the quantum signal and reference pulses using a dynamic polarization controller (DPC), a polarization beam splitter (PBS), a beam splitter (BS), and a photodetector (PD). To measure the quadratures of the quantum signal and reference pulses, a continuous laser with a center wavelength of $1550.12$ nm was used as LLO instead of the traditional pulsed LLO, which requires a high common-mode rejection ratio that is difficult to achieve with a short pulse.

During the quantum signal transmission, the laser pulses were sent in frame, with each frame containing $10^7$ pulses. All output signals of the $\text{HDs}$ were sampled using OSs, with $12$-bit $50$-GSample/s analog-to-digital converters. In the offline digital signal processing, a single quadrature was obtained by integrating the respective pulsed signal and normalizing to the shot noise~\cite{Chi:10}. More details about heterodyne detection can be seen in Section I of the supplemental material~\cite{Supplemental}. 

%\section{Phase rotation and discretization}

\textit{Phase rotation and discretization}---In heterodyne detection, the relative phase of each laser pulse can be determined by the measured quadratures $X$ and $P$. Three sets of relative phases are measured in our experiment
\begin{equation}
	\begin{aligned}
	\text{HD}_1: \Phi_1 &= \varphi_{\text{S}} - \varphi_{\text{R}} + \theta_{\text{S1}}, \\
	\text{HD}_2: \Phi_2 &= \varphi_{\text{R}} - \varphi_{\text{L}} + \theta_{\text{F2}} + \theta_{\text{S2}}, \\
	\text{HD}_3: \Phi_3 &= \varphi_{\text{S}} - \varphi_{\text{L}} + \theta_{\text{F3}} + \theta_{\text{S3}} + \Delta\theta,
	\end{aligned}
\end{equation}
where $\varphi_{\text{S}}$, $\varphi_{\text{R}}$, and $\varphi_{\text{L}}$ are the initial phases of the signal, reference, and LLO beams, respectively, and $\varphi_{\text{S}}$, and $\varphi_{\text{R}}$ are randomly distributed for each pulse. As the linewidth of the LLO beam is of the order of megahertz, and the laser pulse width is 200 ps, $\varphi_\text{L}$ can be treated as a constant value within each pulse. The slow-drift phases $\theta_{\text{S1}}$, $\theta_{\text{S2}}$, and $\theta_{\text{S3}}$ arise from the optical path difference of the several meters fiber, and the corresponding drift speed is several rad/s. The fast-drift phases $\theta_{\text{F2}}$ and $\theta_{\text{F3}}$, which are caused by the same long single-mode fiber (quantum channel), are equal and the drift speeds are the order of krad/s at $100$ km. Due to low SNR for the quantum signal, the effect of the shot noise on the relative phase $\Phi_3$ cannot be neglected and denoted as $\Delta\theta$.

To match the relative phases between Alice and Bob, Bob rotates $\Phi_3$ according to $\Phi_2$ to eliminate the relative phases $\varphi_{\text{L}}$, $\theta_{\text{F2}}$ and $\theta_{\text{F3}}$. This rotation yields the relative phase
\begin{equation}
\Phi_{\text{R3}} = \Phi_3 - \Phi_2 = \varphi_{\text{S}} - \varphi_{\text{R}} + \theta_{{\text{S32}}} + \Delta\theta,
\end{equation}
where $\theta_{\text{S32}} = \theta_{\text{F3}} - \theta_{\text{F2}}$. Next, they split each frame into 100 blocks; in addition, Alice declares her phases $\Phi_1$ for part of each block (10$\%$). Then, Bob searches for the slow-drift phases $\theta_{\text{max}}$ by calculating the correlation among the selected data samples; the phase is related to $\theta_{\text{S1}}$ and $\theta_{\text{S32}}$ for each block:
\begin{equation}
\theta_{\text{S1}} = \theta_{\text{S32}} + \theta_{\text{max}}.
\end{equation}

By using $\theta_{\text{max}}$, Bob compensate the slow-drift phase in each block and get
\begin{equation}
	\Phi_{\text{M3}}=\Phi_{\text{R3}}+\theta_{\text{max}}=\Phi_1+\Delta\theta. 
\end{equation}
Notably, all the phases are defined in the range $[0, 2\pi)$. For more details on the phase drift, please refer to Section II of the supplemental material~\cite{Supplemental}.

After phase compensation, Alice transforms the remained $\Phi_1$ using the following criteria and announces publicly the phases $\Delta \Phi_1$
\begin{align}
	\left\lfloor \Phi_1, \frac{\pi}{2} \right\rfloor = k, ~k = (0, 1, 2, 3), 
	~\text{mod} \left( \Phi_1, \frac{\pi}{2} \right) = \Delta \Phi_1.
	\label{eq:phase_transform} 
\end{align}
Subsequently, Bob rotates $\Phi_{\text{M3}}$ according to $\Delta \Phi_1$ and get $\phi_k=\Phi_{\text{M3}}-\Delta \Phi_1$. He then obtained the final measurement result $y = |y| e^{i\phi_k} \in \mathbb{C}$ and his raw key string via the following key map:
\begin{equation}
k' = j, ~\text{if } \phi_k \in \left[ \frac{(2j-1)\pi}{4}, \frac{(2j+1)\pi}{4} \right), 
\end{equation}
where $j= (0, 1, 2, 3)$. At this stage, Alice and Bob obtain raw key string $(k, k')$.
%\section{Experimental results}
\begin{table}[b]
	\makeatletter
	\renewcommand{\@makecaption}[2]{%
		\vskip\abovecaptionskip
		\raggedright #1. #2\par 
		\vskip\belowcaptionskip}
	\makeatother
	\centering
	%\captionsetup{justification=raggedright, singlelinecheck=false}
	\caption{Experimental parameters.}
	\label{tab:1}
	\begin{ruledtabular}
		\begin{tabular}{ccccccc}
			\multicolumn{4}{c}{$|\alpha|~(\sqrt{\text{SNU}})$} & \multirow{2}{*}{$V_{el}~(\text{SNU})$} & \multirow{2}{*}{$\eta$} & \multirow{2}{*}{$\beta$} \\
			\cmidrule(lr){1-4}
			2\,\si{m} & 25\,\si{km} & 50\,\si{km} & 100\,\si{km} & & & \\
			\midrule
			1.15 & 0.821 & 0.773 & 0.753 & 0.189 & 0.352 & 0.95 \\
		\end{tabular}
	\end{ruledtabular}
\end{table}

\begin{figure}
	\includegraphics[width=\columnwidth]{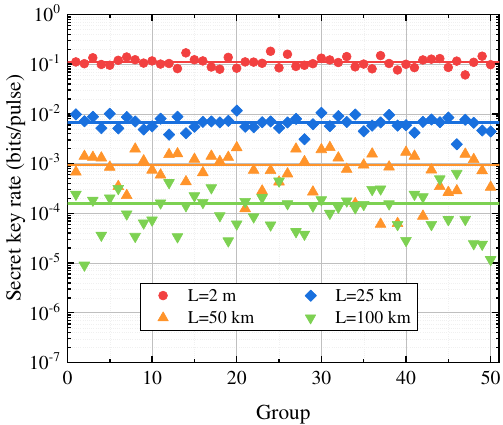}
	\caption{The secret key rates at different transmission lengths: 2 m (red dots), 25 km (blue diamonds), 50 km (orange upright triangles), and 100 km (green inverted triangles). The lines represent the average values of corresponding secret key rates.}
	\label{fig:3}
\end{figure}

\textit{Experimental results}---The secret key rates $R$ at four different transmission lengths (2 m, 25 km, 50 km, and 100 km) are presented in Fig.~\ref{fig:3}, where the relevant parameters including the modulation amplitude $|\alpha|$, electronic noise $V_{el}$, detection efficiency $\eta$, and reverse reconciliation efficiency $\beta$ are listed in Table~\ref{tab:1}. To characterize the stability of our system, the secret key rates of 50 groups of pulses were recorded (different shapes) for each transmission length, and the corresponding average values were also drawn (solid lines).

The success rate $\eta_{\text{s}}$ of estimating the secret key rate $R$ is sensitive to the fluctuations of the first- and second-order moments of the quadratures imposed by the finite-size effect. The amount of quantum signal pulses $N$ in a group and success rate $\eta_{\text{s}}$ at different transmission lengths $L$ are listed in Table~\ref{tab:2}. When the transmission length is longer, greater $N$ is required to maintain a sufficiently high $\eta_{\text{s}}$. It means that as the transmission length increases, $R$ becomes more sensitive to the statistical fluctuation of the first- and second-order moments of the quadratures. To suppress the finite-size effect, more data are required. 
\begin{table}[htbp]
	\makeatletter
	\renewcommand{\@makecaption}[2]{%
		\vskip\abovecaptionskip
		\raggedright #1. #2\par 
		\vskip\belowcaptionskip}
	\makeatother
	\centering
	\caption{Results for different transmission lengths.}
	\label{tab:2}
	\begin{ruledtabular}
		\begin{tabular}{cccccc}
			\makecell{$L$\\$\text{(km)}$} & $N$ & $\eta_{\text{s}}$ & \makecell{$\langle R \rangle$\\$\text{(bits/pulse)}$} & \makecell{$\varepsilon_{\text{e}}$ \\ $\text{(SNU)}$} & \makecell{$K$\\ $\text{(Mbps)}$} \\ \hline
			0.002 & $10^{8}$ & 100\% & $1.11\times10^{-1}$ & 0.0239 & 89.9 \\ \hline
			\multirow{2}{*}{25} & $10^{8}$ & 90\% & \multirow{2}{*}{$6.73\times10^{-3}$} & \multirow{2}{*}{0.0281} & \multirow{2}{*}{5.45} \\ \cmidrule(lr){2-3}
			& $10^{9}$ & 100\% &  &  & \\ \hline
			\multirow{2}{*}{50} & $10^{9}$ & 92\% & \multirow{2}{*}{$9.33\times10^{-4}$} & \multirow{2}{*}{0.0291} & \multirow{2}{*}{0.756} \\ \cmidrule(lr){2-3}
			& $10^{10}$ & 96\% &  &  & \\ \hline
			100 & $10^{10}$ & 82\% & $1.57\times10^{-4}$ & 0.0262 & 0.127 \\
		\end{tabular}
	\end{ruledtabular}
\end{table}

To evaluate the performance of our experimental system, the equivalent excess noises $\varepsilon_{\text{e}}$ are listed. To this end, a linear quantum channel is assumed and the secret key rate is set to the experimental average secret key rate. By using the experimental parameters in Table~\ref{tab:1}, the equivalent excess noise $\varepsilon_{\text{e}}$ can be evaluated. To estimate the secret key bit rate $K$, frame error rate of $10\%$ in reverse reconciliation is assumed.

Our system achieves the maximum transmission length as that of the Gaussian modulation LLO CV-QKD protocol that reported in Ref.~\cite{Hajomer2024}, and our secret key bit rate exceeds its results by five times. The secret key rate is one order of magnitude higher than that of the passive DV protocols~\cite{Lu:23,Hu:23}, and covers a longer transmission length. More detailed performance comparison between our system and the existing systems can be seen in Section III of the supplemental material~\cite{Supplemental}.

%\section{CONCLUSIONS}

\textit{Conclusions and outlook}---We propose and demonstrate a high-performance fully passive discrete-state LLO CV-QKD protocol that eliminates all modulator side channels on the source side. To this end, three key technologies are employed including independently seeded, gain-switched pulsed laser sources, interference of continuous laser beam and short pulsed laser beam, and the phase rotation and discretization approach. Secure key distribution is achieved over $100$ km low-loss single-mode fiber with a repetition rate of $1$ GHz. Our protocol significantly simplifies the architecture of CV-QKD system by eliminating the need of optical modulators and random number generators, while presents robust practical security and superior performance.

In the future, we will increase the system repetition rate to above 10 GHz considering that pulse width can be reduced to $\sim 20$ $\text{ps}$~\cite{Comandar:16}. Furthermore, the present four-state protocol can be extended to eight or more discrete states protocol to improve the system performance, and finite-size security will also be investigated~\cite{Kanitschar:23}. 

\textit{Acknowledgments}---This work was supported in part by the National Natural Science Foundation of China under Grants 11504219, 62175138, 62205188, and 11904219, in part by the Innovation Program for Quantum Science and Technology under Grant 2021ZD0300703, and in part by the Shanxi Provincial Foundation for Returned Scholars, China, under Grant 2022-016.

\bibliography{Manuscript}     % Produces the bibliography via BibTeX.

\providecommand{\noopsort}[1]{}
\begin{thebibliography}{70}%
\makeatletter
\providecommand \@ifxundefined [1]{%
 \@ifx{#1\undefined}
}%
\providecommand \@ifnum [1]{%
 \ifnum #1\expandafter \@firstoftwo
 \else \expandafter \@secondoftwo
 \fi
}%
\providecommand \@ifx [1]{%
 \ifx #1\expandafter \@firstoftwo
 \else \expandafter \@secondoftwo
 \fi
}%
\providecommand \natexlab [1]{#1}%
\providecommand \enquote  [1]{``#1''}%
\providecommand \bibnamefont  [1]{#1}%
\providecommand \bibfnamefont [1]{#1}%
\providecommand \citenamefont [1]{#1}%
\providecommand \href@noop [0]{\@secondoftwo}%
\providecommand \href [0]{\begingroup \@sanitize@url \@href}%
\providecommand \@href[1]{\@@startlink{#1}\@@href}%
\providecommand \@@href[1]{\endgroup#1\@@endlink}%
\providecommand \@sanitize@url [0]{\catcode `\\12\catcode `\$12\catcode
  `\&12\catcode `\#12\catcode `\^12\catcode `\_12\catcode `\%12\relax}%
\providecommand \@@startlink[1]{}%
\providecommand \@@endlink[0]{}%
\providecommand \url  [0]{\begingroup\@sanitize@url \@url }%
\providecommand \@url [1]{\endgroup\@href {#1}{\urlprefix }}%
\providecommand \urlprefix  [0]{URL }%
\providecommand \Eprint [0]{\href }%
\providecommand \doibase [0]{http://dx.doi.org/}%
\providecommand \selectlanguage [0]{\@gobble}%
\providecommand \bibinfo  [0]{\@secondoftwo}%
\providecommand \bibfield  [0]{\@secondoftwo}%
\providecommand \translation [1]{[#1]}%
\providecommand \BibitemOpen [0]{}%
\providecommand \bibitemStop [0]{}%
\providecommand \bibitemNoStop [0]{.\EOS\space}%
\providecommand \EOS [0]{\spacefactor3000\relax}%
\providecommand \BibitemShut  [1]{\csname bibitem#1\endcsname}%
\let\auto@bib@innerbib\@empty
%</preamble>
\bibitem [{\citenamefont {Xu}\ \emph {et~al.}(2020)\citenamefont {Xu},
  \citenamefont {Ma}, \citenamefont {Zhang}, \citenamefont {Lo},\ and\
  \citenamefont {Pan}}]{RevModPhys.92.025002}%
  \BibitemOpen
  \bibfield  {author} {\bibinfo {author} {\bibfnamefont {F.}~\bibnamefont
  {Xu}}, \bibinfo {author} {\bibfnamefont {X.}~\bibnamefont {Ma}}, \bibinfo
  {author} {\bibfnamefont {Q.}~\bibnamefont {Zhang}}, \bibinfo {author}
  {\bibfnamefont {H.-K.}\ \bibnamefont {Lo}}, \ and\ \bibinfo {author}
  {\bibfnamefont {J.-W.}\ \bibnamefont {Pan}},\ }\href {\doibase
  10.1103/RevModPhys.92.025002} {\bibfield  {journal} {\bibinfo  {journal}
  {Rev. Mod. Phys.}\ }\textbf {\bibinfo {volume} {92}},\ \bibinfo {pages}
  {025002} (\bibinfo {year} {2020})}\BibitemShut {NoStop}%
\bibitem [{\citenamefont {Zhang}\ \emph {et~al.}(2024)\citenamefont {Zhang},
  \citenamefont {Bian}, \citenamefont {Li}, \citenamefont {Yu},\ and\
  \citenamefont {Guo}}]{Zhang2024}%
  \BibitemOpen
  \bibfield  {author} {\bibinfo {author} {\bibfnamefont {Y.~C.}\ \bibnamefont
  {Zhang}}, \bibinfo {author} {\bibfnamefont {Y.~M.}\ \bibnamefont {Bian}},
  \bibinfo {author} {\bibfnamefont {Z.~Y.}\ \bibnamefont {Li}}, \bibinfo
  {author} {\bibfnamefont {S.}~\bibnamefont {Yu}}, \ and\ \bibinfo {author}
  {\bibfnamefont {H.}~\bibnamefont {Guo}},\ }\href {\doibase 10.1063/5.0179566}
  {\bibfield  {journal} {\bibinfo  {journal} {Appl. Phys. Rev.}\ }\textbf
  {\bibinfo {volume} {11}},\ \bibinfo {pages} {011304} (\bibinfo {year}
  {2024})}\BibitemShut {NoStop}%
\bibitem [{\citenamefont {Pirandola}\ \emph {et~al.}(2020)\citenamefont
  {Pirandola}, \citenamefont {Andersen}, \citenamefont {Banchi}, \citenamefont
  {Berta}, \citenamefont {Bunandar}, \citenamefont {Colbeck}, \citenamefont
  {Englund}, \citenamefont {Gehring}, \citenamefont {Lupo}, \citenamefont
  {Ottaviani},\ and\ \citenamefont {et~al.}}]{Pirandola:20}%
  \BibitemOpen
  \bibfield  {author} {\bibinfo {author} {\bibfnamefont {S.}~\bibnamefont
  {Pirandola}}, \bibinfo {author} {\bibfnamefont {U.~L.}\ \bibnamefont
  {Andersen}}, \bibinfo {author} {\bibfnamefont {L.}~\bibnamefont {Banchi}},
  \bibinfo {author} {\bibfnamefont {M.}~\bibnamefont {Berta}}, \bibinfo
  {author} {\bibfnamefont {D.}~\bibnamefont {Bunandar}}, \bibinfo {author}
  {\bibfnamefont {R.}~\bibnamefont {Colbeck}}, \bibinfo {author} {\bibfnamefont
  {D.}~\bibnamefont {Englund}}, \bibinfo {author} {\bibfnamefont
  {T.}~\bibnamefont {Gehring}}, \bibinfo {author} {\bibfnamefont
  {C.}~\bibnamefont {Lupo}}, \bibinfo {author} {\bibfnamefont {C.}~\bibnamefont
  {Ottaviani}}, \ and\ \bibinfo {author} {\bibnamefont {et~al.}},\ }\href
  {\doibase 10.1364/AOP.361502} {\bibfield  {journal} {\bibinfo  {journal}
  {Adv. Opt. Photon.}\ }\textbf {\bibinfo {volume} {12}},\ \bibinfo {pages}
  {1012} (\bibinfo {year} {2020})}\BibitemShut {NoStop}%
\bibitem [{\citenamefont {Weedbrook}\ \emph {et~al.}(2012)\citenamefont
  {Weedbrook}, \citenamefont {Pirandola}, \citenamefont
  {Garc{\'{\i}}a-Patr{\'o}n}, \citenamefont {Cerf}, \citenamefont {Ralph},
  \citenamefont {Shapiro},\ and\ \citenamefont {Lloyd}}]{RevModPhys.84.621}%
  \BibitemOpen
  \bibfield  {author} {\bibinfo {author} {\bibfnamefont {C.}~\bibnamefont
  {Weedbrook}}, \bibinfo {author} {\bibfnamefont {S.}~\bibnamefont
  {Pirandola}}, \bibinfo {author} {\bibfnamefont {R.}~\bibnamefont
  {Garc{\'{\i}}a-Patr{\'o}n}}, \bibinfo {author} {\bibfnamefont {N.~J.}\
  \bibnamefont {Cerf}}, \bibinfo {author} {\bibfnamefont {T.~C.}\ \bibnamefont
  {Ralph}}, \bibinfo {author} {\bibfnamefont {J.~H.}\ \bibnamefont {Shapiro}},
  \ and\ \bibinfo {author} {\bibfnamefont {S.}~\bibnamefont {Lloyd}},\ }\href
  {\doibase 10.1103/RevModPhys.84.621} {\bibfield  {journal} {\bibinfo
  {journal} {Rev. Mod. Phys.}\ }\textbf {\bibinfo {volume} {84}},\ \bibinfo
  {pages} {621} (\bibinfo {year} {2012})}\BibitemShut {NoStop}%
\bibitem [{\citenamefont {Scarani}\ \emph {et~al.}(2009)\citenamefont
  {Scarani}, \citenamefont {Bechmann-Pasquinucci}, \citenamefont {Cerf},
  \citenamefont {Du{\v s}ek}, \citenamefont {L{\" u}tkenhaus},\ and\
  \citenamefont {Peev}}]{RevModPhys.81.1301}%
  \BibitemOpen
  \bibfield  {author} {\bibinfo {author} {\bibfnamefont {V.}~\bibnamefont
  {Scarani}}, \bibinfo {author} {\bibfnamefont {H.}~\bibnamefont
  {Bechmann-Pasquinucci}}, \bibinfo {author} {\bibfnamefont {N.~J.}\
  \bibnamefont {Cerf}}, \bibinfo {author} {\bibfnamefont {M.}~\bibnamefont
  {Du{\v s}ek}}, \bibinfo {author} {\bibfnamefont {N.}~\bibnamefont {L{\"
  u}tkenhaus}}, \ and\ \bibinfo {author} {\bibfnamefont {M.}~\bibnamefont
  {Peev}},\ }\href {\doibase 10.1103/RevModPhys.81.1301} {\bibfield  {journal}
  {\bibinfo  {journal} {Rev. Mod. Phys.}\ }\textbf {\bibinfo {volume} {81}},\
  \bibinfo {pages} {1301} (\bibinfo {year} {2009})}\BibitemShut {NoStop}%
\bibitem [{\citenamefont {Liao}\ \emph {et~al.}(2017)\citenamefont {Liao},
  \citenamefont {Cai}, \citenamefont {Liu}, \citenamefont {Zhang},
  \citenamefont {Li}, \citenamefont {Ren}, \citenamefont {Yin}, \citenamefont
  {Shen}, \citenamefont {Cao}, \citenamefont {Li},\ and\ \citenamefont
  {et~al.}}]{Liao2017}%
  \BibitemOpen
  \bibfield  {author} {\bibinfo {author} {\bibfnamefont {S.~K.}\ \bibnamefont
  {Liao}}, \bibinfo {author} {\bibfnamefont {W.~Q.}\ \bibnamefont {Cai}},
  \bibinfo {author} {\bibfnamefont {W.~Y.}\ \bibnamefont {Liu}}, \bibinfo
  {author} {\bibfnamefont {L.}~\bibnamefont {Zhang}}, \bibinfo {author}
  {\bibfnamefont {Y.}~\bibnamefont {Li}}, \bibinfo {author} {\bibfnamefont
  {J.~G.}\ \bibnamefont {Ren}}, \bibinfo {author} {\bibfnamefont
  {J.}~\bibnamefont {Yin}}, \bibinfo {author} {\bibfnamefont {Q.}~\bibnamefont
  {Shen}}, \bibinfo {author} {\bibfnamefont {Y.}~\bibnamefont {Cao}}, \bibinfo
  {author} {\bibfnamefont {Z.~P.}\ \bibnamefont {Li}}, \ and\ \bibinfo {author}
  {\bibnamefont {et~al.}},\ }\href {\doibase 10.1038/nature23655} {\bibfield
  {journal} {\bibinfo  {journal} {Nature}\ }\textbf {\bibinfo {volume} {549}},\
  \bibinfo {pages} {43} (\bibinfo {year} {2017})}\BibitemShut {NoStop}%
\bibitem [{\citenamefont {Liu}\ \emph {et~al.}(2023)\citenamefont {Liu},
  \citenamefont {Zhang}, \citenamefont {Jiang}, \citenamefont {Chen},
  \citenamefont {Zhang}, \citenamefont {Pan}, \citenamefont {Ma}, \citenamefont
  {Dong}, \citenamefont {Xiong}, \citenamefont {Zhang},\ and\ \citenamefont
  {et~al.}}]{PhysRevLett.130.210801}%
  \BibitemOpen
  \bibfield  {author} {\bibinfo {author} {\bibfnamefont {Y.}~\bibnamefont
  {Liu}}, \bibinfo {author} {\bibfnamefont {W.~J.}\ \bibnamefont {Zhang}},
  \bibinfo {author} {\bibfnamefont {C.}~\bibnamefont {Jiang}}, \bibinfo
  {author} {\bibfnamefont {J.~P.}\ \bibnamefont {Chen}}, \bibinfo {author}
  {\bibfnamefont {C.}~\bibnamefont {Zhang}}, \bibinfo {author} {\bibfnamefont
  {W.~X.}\ \bibnamefont {Pan}}, \bibinfo {author} {\bibfnamefont
  {D.}~\bibnamefont {Ma}}, \bibinfo {author} {\bibfnamefont {H.}~\bibnamefont
  {Dong}}, \bibinfo {author} {\bibfnamefont {J.~M.}\ \bibnamefont {Xiong}},
  \bibinfo {author} {\bibfnamefont {C.~J.}\ \bibnamefont {Zhang}}, \ and\
  \bibinfo {author} {\bibnamefont {et~al.}},\ }\href {\doibase
  10.1103/PhysRevLett.130.210801} {\bibfield  {journal} {\bibinfo  {journal}
  {Phys. Rev. Lett.}\ }\textbf {\bibinfo {volume} {130}},\ \bibinfo {pages}
  {210801} (\bibinfo {year} {2023})}\BibitemShut {NoStop}%
\bibitem [{\citenamefont {Zhang}\ \emph {et~al.}(2020)\citenamefont {Zhang},
  \citenamefont {Chen}, \citenamefont {Pirandola}, \citenamefont {Wang},
  \citenamefont {Zhou}, \citenamefont {Chu}, \citenamefont {Zhao},
  \citenamefont {Xu}, \citenamefont {Yu},\ and\ \citenamefont
  {Guo}}]{PhysRevLett.125.010502}%
  \BibitemOpen
  \bibfield  {author} {\bibinfo {author} {\bibfnamefont {Y.~C.}\ \bibnamefont
  {Zhang}}, \bibinfo {author} {\bibfnamefont {Z.~Y.}\ \bibnamefont {Chen}},
  \bibinfo {author} {\bibfnamefont {S.}~\bibnamefont {Pirandola}}, \bibinfo
  {author} {\bibfnamefont {X.~Y.}\ \bibnamefont {Wang}}, \bibinfo {author}
  {\bibfnamefont {C.}~\bibnamefont {Zhou}}, \bibinfo {author} {\bibfnamefont
  {B.~J.}\ \bibnamefont {Chu}}, \bibinfo {author} {\bibfnamefont {Y.~J.}\
  \bibnamefont {Zhao}}, \bibinfo {author} {\bibfnamefont {B.~J.}\ \bibnamefont
  {Xu}}, \bibinfo {author} {\bibfnamefont {S.}~\bibnamefont {Yu}}, \ and\
  \bibinfo {author} {\bibfnamefont {H.}~\bibnamefont {Guo}},\ }\href {\doibase
  10.1103/PhysRevLett.125.010502} {\bibfield  {journal} {\bibinfo  {journal}
  {Phys. Rev. Lett.}\ }\textbf {\bibinfo {volume} {125}},\ \bibinfo {pages}
  {010502} (\bibinfo {year} {2020})}\BibitemShut {NoStop}%
\bibitem [{\citenamefont {Hajomer}\ \emph
  {et~al.}(2024{\natexlab{a}})\citenamefont {Hajomer}, \citenamefont {Derkach},
  \citenamefont {Jain}, \citenamefont {Chin}, \citenamefont {Andersen},\ and\
  \citenamefont {Gehring}}]{Hajomer2024}%
  \BibitemOpen
  \bibfield  {author} {\bibinfo {author} {\bibfnamefont {A.~A.~E.}\
  \bibnamefont {Hajomer}}, \bibinfo {author} {\bibfnamefont {I.}~\bibnamefont
  {Derkach}}, \bibinfo {author} {\bibfnamefont {N.}~\bibnamefont {Jain}},
  \bibinfo {author} {\bibfnamefont {H.~M.}\ \bibnamefont {Chin}}, \bibinfo
  {author} {\bibfnamefont {U.~L.}\ \bibnamefont {Andersen}}, \ and\ \bibinfo
  {author} {\bibfnamefont {T.}~\bibnamefont {Gehring}},\ }\href {\doibase
  10.1126/sciadv.adi9474} {\bibfield  {journal} {\bibinfo  {journal} {Sci.
  Adv.}\ }\textbf {\bibinfo {volume} {10}},\ \bibinfo {pages} {eadi9474}
  (\bibinfo {year} {2024}{\natexlab{a}})}\BibitemShut {NoStop}%
\bibitem [{\citenamefont {Zhang}\ \emph {et~al.}(2019)\citenamefont {Zhang},
  \citenamefont {Haw}, \citenamefont {Cai}, \citenamefont {Xu}, \citenamefont
  {Assad}, \citenamefont {Fitzsimons}, \citenamefont {Zhou}, \citenamefont
  {Zhang}, \citenamefont {Yu}, \citenamefont {Wu},\ and\ \citenamefont
  {et~al.}}]{Zhang2019}%
  \BibitemOpen
  \bibfield  {author} {\bibinfo {author} {\bibfnamefont {G.}~\bibnamefont
  {Zhang}}, \bibinfo {author} {\bibfnamefont {J.~Y.}\ \bibnamefont {Haw}},
  \bibinfo {author} {\bibfnamefont {H.}~\bibnamefont {Cai}}, \bibinfo {author}
  {\bibfnamefont {F.}~\bibnamefont {Xu}}, \bibinfo {author} {\bibfnamefont
  {S.~M.}\ \bibnamefont {Assad}}, \bibinfo {author} {\bibfnamefont {J.~F.}\
  \bibnamefont {Fitzsimons}}, \bibinfo {author} {\bibfnamefont
  {X.}~\bibnamefont {Zhou}}, \bibinfo {author} {\bibfnamefont {Y.}~\bibnamefont
  {Zhang}}, \bibinfo {author} {\bibfnamefont {S.}~\bibnamefont {Yu}}, \bibinfo
  {author} {\bibfnamefont {J.}~\bibnamefont {Wu}}, \ and\ \bibinfo {author}
  {\bibnamefont {et~al.}},\ }\href {\doibase 10.1038/s41566-019-0504-5}
  {\bibfield  {journal} {\bibinfo  {journal} {Nat. Photon.}\ }\textbf {\bibinfo
  {volume} {13}},\ \bibinfo {pages} {839} (\bibinfo {year} {2019})}\BibitemShut
  {NoStop}%
\bibitem [{\citenamefont {Li}\ \emph {et~al.}(2023{\natexlab{a}})\citenamefont
  {Li}, \citenamefont {Wang}, \citenamefont {Li}, \citenamefont {Huang},
  \citenamefont {Guo}, \citenamefont {Lu}, \citenamefont {Zhou},\ and\
  \citenamefont {Zeng}}]{Li:18}%
  \BibitemOpen
  \bibfield  {author} {\bibinfo {author} {\bibfnamefont {L.}~\bibnamefont
  {Li}}, \bibinfo {author} {\bibfnamefont {T.}~\bibnamefont {Wang}}, \bibinfo
  {author} {\bibfnamefont {X.~H.}\ \bibnamefont {Li}}, \bibinfo {author}
  {\bibfnamefont {P.}~\bibnamefont {Huang}}, \bibinfo {author} {\bibfnamefont
  {Y.~Y.}\ \bibnamefont {Guo}}, \bibinfo {author} {\bibfnamefont {L.~J.}\
  \bibnamefont {Lu}}, \bibinfo {author} {\bibfnamefont {L.~J.}\ \bibnamefont
  {Zhou}}, \ and\ \bibinfo {author} {\bibfnamefont {G.~H.}\ \bibnamefont
  {Zeng}},\ }\href {\doibase 10.1364/PRJ.474457} {\bibfield  {journal}
  {\bibinfo  {journal} {Photonics Res.}\ }\textbf {\bibinfo {volume} {11}},\
  \bibinfo {pages} {504} (\bibinfo {year} {2023}{\natexlab{a}})}\BibitemShut
  {NoStop}%
\bibitem [{\citenamefont {Jia}\ \emph {et~al.}(2023)\citenamefont {Jia},
  \citenamefont {Wang}, \citenamefont {Hu}, \citenamefont {Hua}, \citenamefont
  {Zhang}, \citenamefont {Guo}, \citenamefont {Zhang}, \citenamefont {Xiao},
  \citenamefont {Yu}, \citenamefont {Zou},\ and\ \citenamefont
  {et~al}}]{Jia_2023}%
  \BibitemOpen
  \bibfield  {author} {\bibinfo {author} {\bibfnamefont {Y.~X.}\ \bibnamefont
  {Jia}}, \bibinfo {author} {\bibfnamefont {X.~Y.}\ \bibnamefont {Wang}},
  \bibinfo {author} {\bibfnamefont {X.}~\bibnamefont {Hu}}, \bibinfo {author}
  {\bibfnamefont {X.}~\bibnamefont {Hua}}, \bibinfo {author} {\bibfnamefont
  {Y.}~\bibnamefont {Zhang}}, \bibinfo {author} {\bibfnamefont {X.~B.}\
  \bibnamefont {Guo}}, \bibinfo {author} {\bibfnamefont {S.~X.}\ \bibnamefont
  {Zhang}}, \bibinfo {author} {\bibfnamefont {X.}~\bibnamefont {Xiao}},
  \bibinfo {author} {\bibfnamefont {S.~H.}\ \bibnamefont {Yu}}, \bibinfo
  {author} {\bibfnamefont {J.}~\bibnamefont {Zou}}, \ and\ \bibinfo {author}
  {\bibnamefont {et~al}},\ }\href {\doibase 10.1088/1367-2630/acfcd4}
  {\bibfield  {journal} {\bibinfo  {journal} {New J. Phys.}\ }\textbf {\bibinfo
  {volume} {25}},\ \bibinfo {pages} {103030} (\bibinfo {year}
  {2023})}\BibitemShut {NoStop}%
\bibitem [{\citenamefont {Bian}\ \emph {et~al.}(2024)\citenamefont {Bian},
  \citenamefont {Pan}, \citenamefont {Xu}, \citenamefont {Zhao}, \citenamefont
  {Li}, \citenamefont {Huang}, \citenamefont {Zhang}, \citenamefont {Yu},
  \citenamefont {Zhang},\ and\ \citenamefont {Xu}}]{Bian_2024}%
  \BibitemOpen
  \bibfield  {author} {\bibinfo {author} {\bibfnamefont {Y.~M.}\ \bibnamefont
  {Bian}}, \bibinfo {author} {\bibfnamefont {Y.}~\bibnamefont {Pan}}, \bibinfo
  {author} {\bibfnamefont {X.~S.}\ \bibnamefont {Xu}}, \bibinfo {author}
  {\bibfnamefont {L.}~\bibnamefont {Zhao}}, \bibinfo {author} {\bibfnamefont
  {Y.}~\bibnamefont {Li}}, \bibinfo {author} {\bibfnamefont {W.}~\bibnamefont
  {Huang}}, \bibinfo {author} {\bibfnamefont {L.}~\bibnamefont {Zhang}},
  \bibinfo {author} {\bibfnamefont {S.}~\bibnamefont {Yu}}, \bibinfo {author}
  {\bibfnamefont {Y.~C.}\ \bibnamefont {Zhang}}, \ and\ \bibinfo {author}
  {\bibfnamefont {B.~J.}\ \bibnamefont {Xu}},\ }\href {\doibase
  10.1063/5.0203130} {\bibfield  {journal} {\bibinfo  {journal} {Appl. Phys.
  Lett.}\ }\textbf {\bibinfo {volume} {124}},\ \bibinfo {pages} {174001}
  (\bibinfo {year} {2024})}\BibitemShut {NoStop}%
\bibitem [{\citenamefont {Piétri}\ \emph {et~al.}(2024)\citenamefont
  {Piétri}, \citenamefont {Vidarte}, \citenamefont {Schiavon}, \citenamefont
  {Vivien}, \citenamefont {Grangier}, \citenamefont {Rhouni},\ and\
  \citenamefont {Diamanti}}]{Pietri:24}%
  \BibitemOpen
  \bibfield  {author} {\bibinfo {author} {\bibfnamefont {Y.}~\bibnamefont
  {Piétri}}, \bibinfo {author} {\bibfnamefont {L.~T.}\ \bibnamefont
  {Vidarte}}, \bibinfo {author} {\bibfnamefont {M.}~\bibnamefont {Schiavon}},
  \bibinfo {author} {\bibfnamefont {L.}~\bibnamefont {Vivien}}, \bibinfo
  {author} {\bibfnamefont {P.}~\bibnamefont {Grangier}}, \bibinfo {author}
  {\bibfnamefont {A.}~\bibnamefont {Rhouni}}, \ and\ \bibinfo {author}
  {\bibfnamefont {E.}~\bibnamefont {Diamanti}},\ }\href {\doibase
  10.1364/OPTICAQ.534699} {\bibfield  {journal} {\bibinfo  {journal} {Opt.
  Quantum}\ }\textbf {\bibinfo {volume} {2}},\ \bibinfo {pages} {428} (\bibinfo
  {year} {2024})}\BibitemShut {NoStop}%
\bibitem [{\citenamefont {Hajomer}\ \emph
  {et~al.}(2024{\natexlab{b}})\citenamefont {Hajomer}, \citenamefont
  {Bruynsteen}, \citenamefont {Derkach}, \citenamefont {Jain}, \citenamefont
  {Bomhals}, \citenamefont {Bastiaens}, \citenamefont {Andersen}, \citenamefont
  {Yin},\ and\ \citenamefont {Gehring}}]{Hajomer:24}%
  \BibitemOpen
  \bibfield  {author} {\bibinfo {author} {\bibfnamefont {A.~A.~E.}\
  \bibnamefont {Hajomer}}, \bibinfo {author} {\bibfnamefont {C.}~\bibnamefont
  {Bruynsteen}}, \bibinfo {author} {\bibfnamefont {I.}~\bibnamefont {Derkach}},
  \bibinfo {author} {\bibfnamefont {N.}~\bibnamefont {Jain}}, \bibinfo {author}
  {\bibfnamefont {A.}~\bibnamefont {Bomhals}}, \bibinfo {author} {\bibfnamefont
  {S.}~\bibnamefont {Bastiaens}}, \bibinfo {author} {\bibfnamefont {U.~L.}\
  \bibnamefont {Andersen}}, \bibinfo {author} {\bibfnamefont {X.}~\bibnamefont
  {Yin}}, \ and\ \bibinfo {author} {\bibfnamefont {T.}~\bibnamefont
  {Gehring}},\ }\href {\doibase 10.1364/OPTICA.530080} {\bibfield  {journal}
  {\bibinfo  {journal} {Optica}\ }\textbf {\bibinfo {volume} {11}},\ \bibinfo
  {pages} {1197} (\bibinfo {year} {2024}{\natexlab{b}})}\BibitemShut {NoStop}%
\bibitem [{\citenamefont {Huang}\ \emph
  {et~al.}(2021{\natexlab{a}})\citenamefont {Huang}, \citenamefont {Shen},
  \citenamefont {Wang}, \citenamefont {Chen}, \citenamefont {Xu}, \citenamefont
  {Yu},\ and\ \citenamefont {Guo}}]{PhysRevApplied.16.064051}%
  \BibitemOpen
  \bibfield  {author} {\bibinfo {author} {\bibfnamefont {Y.~D.}\ \bibnamefont
  {Huang}}, \bibinfo {author} {\bibfnamefont {T.}~\bibnamefont {Shen}},
  \bibinfo {author} {\bibfnamefont {X.~Y.}\ \bibnamefont {Wang}}, \bibinfo
  {author} {\bibfnamefont {Z.~Y.}\ \bibnamefont {Chen}}, \bibinfo {author}
  {\bibfnamefont {B.~J.}\ \bibnamefont {Xu}}, \bibinfo {author} {\bibfnamefont
  {S.}~\bibnamefont {Yu}}, \ and\ \bibinfo {author} {\bibfnamefont
  {H.}~\bibnamefont {Guo}},\ }\href {\doibase 10.1103/PhysRevApplied.16.064051}
  {\bibfield  {journal} {\bibinfo  {journal} {Phys. Rev. Appl.}\ }\textbf
  {\bibinfo {volume} {16}},\ \bibinfo {pages} {064051} (\bibinfo {year}
  {2021}{\natexlab{a}})}\BibitemShut {NoStop}%
\bibitem [{\citenamefont {Wang}\ \emph
  {et~al.}(2023{\natexlab{a}})\citenamefont {Wang}, \citenamefont {Chen},
  \citenamefont {Li}, \citenamefont {Qi}, \citenamefont {Yu},\ and\
  \citenamefont {Guo}}]{Wang:23}%
  \BibitemOpen
  \bibfield  {author} {\bibinfo {author} {\bibfnamefont {X.~Y.}\ \bibnamefont
  {Wang}}, \bibinfo {author} {\bibfnamefont {Z.~Y.}\ \bibnamefont {Chen}},
  \bibinfo {author} {\bibfnamefont {Z.~H.}\ \bibnamefont {Li}}, \bibinfo
  {author} {\bibfnamefont {D.~K.}\ \bibnamefont {Qi}}, \bibinfo {author}
  {\bibfnamefont {S.}~\bibnamefont {Yu}}, \ and\ \bibinfo {author}
  {\bibfnamefont {H.}~\bibnamefont {Guo}},\ }\href {\doibase 10.1364/OL.487582}
  {\bibfield  {journal} {\bibinfo  {journal} {Opt. Lett.}\ }\textbf {\bibinfo
  {volume} {48}},\ \bibinfo {pages} {3327} (\bibinfo {year}
  {2023}{\natexlab{a}})}\BibitemShut {NoStop}%
\bibitem [{\citenamefont {Ji}\ \emph {et~al.}(2024)\citenamefont {Ji},
  \citenamefont {Huang}, \citenamefont {Wang}, \citenamefont {Jiang},\ and\
  \citenamefont {Zeng}}]{Li:24}%
  \BibitemOpen
  \bibfield  {author} {\bibinfo {author} {\bibfnamefont {F.~Y.}\ \bibnamefont
  {Ji}}, \bibinfo {author} {\bibfnamefont {P.}~\bibnamefont {Huang}}, \bibinfo
  {author} {\bibfnamefont {T.}~\bibnamefont {Wang}}, \bibinfo {author}
  {\bibfnamefont {X.~Q.}\ \bibnamefont {Jiang}}, \ and\ \bibinfo {author}
  {\bibfnamefont {G.~H.}\ \bibnamefont {Zeng}},\ }\href {\doibase
  10.1364/PRJ.519909} {\bibfield  {journal} {\bibinfo  {journal} {Photonics
  Res.}\ }\textbf {\bibinfo {volume} {12}},\ \bibinfo {pages} {1485} (\bibinfo
  {year} {2024})}\BibitemShut {NoStop}%
\bibitem [{\citenamefont {Hajomer}\ \emph
  {et~al.}(2024{\natexlab{c}})\citenamefont {Hajomer}, \citenamefont {Derkach},
  \citenamefont {Filip}, \citenamefont {Andersen}, \citenamefont {Usenko},\
  and\ \citenamefont {Gehring}}]{Hajomer:2024}%
  \BibitemOpen
  \bibfield  {author} {\bibinfo {author} {\bibfnamefont {A.~A.~E.}\
  \bibnamefont {Hajomer}}, \bibinfo {author} {\bibfnamefont {I.}~\bibnamefont
  {Derkach}}, \bibinfo {author} {\bibfnamefont {R.}~\bibnamefont {Filip}},
  \bibinfo {author} {\bibfnamefont {U.~L.}\ \bibnamefont {Andersen}}, \bibinfo
  {author} {\bibfnamefont {V.~C.}\ \bibnamefont {Usenko}}, \ and\ \bibinfo
  {author} {\bibfnamefont {T.}~\bibnamefont {Gehring}},\ }\href {\doibase
  10.1038/s41377-024-01633-9} {\bibfield  {journal} {\bibinfo  {journal} {Light
  Sci. Appl.}\ }\textbf {\bibinfo {volume} {13}},\ \bibinfo {pages} {291}
  (\bibinfo {year} {2024}{\natexlab{c}})}\BibitemShut {NoStop}%
\bibitem [{\citenamefont {Zhang}\ \emph {et~al.}(2022)\citenamefont {Zhang},
  \citenamefont {van Leent}, \citenamefont {Redeker}, \citenamefont {Garthoff},
  \citenamefont {Schwonnek}, \citenamefont {Fertig}, \citenamefont {Eppelt},
  \citenamefont {Rosenfeld}, \citenamefont {Scarani}, \citenamefont {Lim},\
  and\ \citenamefont {et~al.}}]{Zhang:22}%
  \BibitemOpen
  \bibfield  {author} {\bibinfo {author} {\bibfnamefont {W.}~\bibnamefont
  {Zhang}}, \bibinfo {author} {\bibfnamefont {T.}~\bibnamefont {van Leent}},
  \bibinfo {author} {\bibfnamefont {K.}~\bibnamefont {Redeker}}, \bibinfo
  {author} {\bibfnamefont {R.}~\bibnamefont {Garthoff}}, \bibinfo {author}
  {\bibfnamefont {R.}~\bibnamefont {Schwonnek}}, \bibinfo {author}
  {\bibfnamefont {F.}~\bibnamefont {Fertig}}, \bibinfo {author} {\bibfnamefont
  {S.}~\bibnamefont {Eppelt}}, \bibinfo {author} {\bibfnamefont
  {W.}~\bibnamefont {Rosenfeld}}, \bibinfo {author} {\bibfnamefont
  {V.}~\bibnamefont {Scarani}}, \bibinfo {author} {\bibfnamefont {C.~C.-W.}\
  \bibnamefont {Lim}}, \ and\ \bibinfo {author} {\bibnamefont {et~al.}},\
  }\href {\doibase 10.1038/s41586-022-04891-y} {\bibfield  {journal} {\bibinfo
  {journal} {Nature}\ }\textbf {\bibinfo {volume} {607}},\ \bibinfo {pages}
  {687} (\bibinfo {year} {2022})}\BibitemShut {NoStop}%
\bibitem [{\citenamefont {Wooltorton}\ \emph {et~al.}(2024)\citenamefont
  {Wooltorton}, \citenamefont {Brown},\ and\ \citenamefont
  {Colbeck}}]{Wooltorton:24}%
  \BibitemOpen
  \bibfield  {author} {\bibinfo {author} {\bibfnamefont {L.}~\bibnamefont
  {Wooltorton}}, \bibinfo {author} {\bibfnamefont {P.}~\bibnamefont {Brown}}, \
  and\ \bibinfo {author} {\bibfnamefont {R.}~\bibnamefont {Colbeck}},\ }\href
  {\doibase 10.1007/s11432-022-3656-4} {\bibfield  {journal} {\bibinfo
  {journal} {Phys. Rev. Lett.}\ }\textbf {\bibinfo {volume} {132}},\ \bibinfo
  {pages} {210802} (\bibinfo {year} {2024})}\BibitemShut {NoStop}%
\bibitem [{\citenamefont {Tan}\ and\ \citenamefont {Wolf}(2024)}]{Tan:24}%
  \BibitemOpen
  \bibfield  {author} {\bibinfo {author} {\bibfnamefont {E.~Y.-Z.}\
  \bibnamefont {Tan}}\ and\ \bibinfo {author} {\bibfnamefont {R.}~\bibnamefont
  {Wolf}},\ }\href {\doibase 10.1007/s11432-022-3656-4} {\bibfield  {journal}
  {\bibinfo  {journal} {Phys. Rev. Lett.}\ }\textbf {\bibinfo {volume} {133}},\
  \bibinfo {pages} {120803} (\bibinfo {year} {2024})}\BibitemShut {NoStop}%
\bibitem [{\citenamefont {Lo}\ \emph {et~al.}(2012)\citenamefont {Lo},
  \citenamefont {Curty},\ and\ \citenamefont {Qi}}]{Lo:12}%
  \BibitemOpen
  \bibfield  {author} {\bibinfo {author} {\bibfnamefont {H.-K.}\ \bibnamefont
  {Lo}}, \bibinfo {author} {\bibfnamefont {M.}~\bibnamefont {Curty}}, \ and\
  \bibinfo {author} {\bibfnamefont {B.}~\bibnamefont {Qi}},\ }\href {\doibase
  10.1103/PhysRevLett.108.130503} {\bibfield  {journal} {\bibinfo  {journal}
  {Phys. Rev. Lett.}\ }\textbf {\bibinfo {volume} {108}},\ \bibinfo {pages}
  {130503} (\bibinfo {year} {2012})}\BibitemShut {NoStop}%
\bibitem [{\citenamefont {Braunstein}\ and\ \citenamefont
  {Pirandola}(2012)}]{Braunstein:12}%
  \BibitemOpen
  \bibfield  {author} {\bibinfo {author} {\bibfnamefont {S.~L.}\ \bibnamefont
  {Braunstein}}\ and\ \bibinfo {author} {\bibfnamefont {S.}~\bibnamefont
  {Pirandola}},\ }\href {\doibase 10.1103/PhysRevLett.108.130502} {\bibfield
  {journal} {\bibinfo  {journal} {Phys. Rev. Lett.}\ }\textbf {\bibinfo
  {volume} {108}},\ \bibinfo {pages} {130502} (\bibinfo {year}
  {2012})}\BibitemShut {NoStop}%
\bibitem [{\citenamefont {L.}\ \emph {et~al.}(2013)\citenamefont {L.},
  \citenamefont {Chen}, \citenamefont {Wang}, \citenamefont {Liang},
  \citenamefont {Shentu}, \citenamefont {Wang}, \citenamefont {Cui},
  \citenamefont {Yin}, \citenamefont {Liu}, \citenamefont {Li},\ and\
  \citenamefont {et~al.}}]{Liu:13}%
  \BibitemOpen
  \bibfield  {author} {\bibinfo {author} {\bibfnamefont {Y.}~\bibnamefont
  {L.}}, \bibinfo {author} {\bibfnamefont {T.~Y.}\ \bibnamefont {Chen}},
  \bibinfo {author} {\bibfnamefont {L.~J.}\ \bibnamefont {Wang}}, \bibinfo
  {author} {\bibfnamefont {H.}~\bibnamefont {Liang}}, \bibinfo {author}
  {\bibfnamefont {G.~L.}\ \bibnamefont {Shentu}}, \bibinfo {author}
  {\bibfnamefont {J.}~\bibnamefont {Wang}}, \bibinfo {author} {\bibfnamefont
  {K.}~\bibnamefont {Cui}}, \bibinfo {author} {\bibfnamefont {H.~L.}\
  \bibnamefont {Yin}}, \bibinfo {author} {\bibfnamefont {N.~L.}\ \bibnamefont
  {Liu}}, \bibinfo {author} {\bibfnamefont {L.}~\bibnamefont {Li}}, \ and\
  \bibinfo {author} {\bibnamefont {et~al.}},\ }\href {\doibase
  10.1103/PhysRevLett.111.130502} {\bibfield  {journal} {\bibinfo  {journal}
  {Phys. Rev. Lett.}\ }\textbf {\bibinfo {volume} {111}},\ \bibinfo {pages}
  {130502} (\bibinfo {year} {2013})}\BibitemShut {NoStop}%
\bibitem [{\citenamefont {Yin}\ \emph {et~al.}(2016)\citenamefont {Yin},
  \citenamefont {Chen}, \citenamefont {Yu}, \citenamefont {Liu}, \citenamefont
  {You}, \citenamefont {Zhou}, \citenamefont {Chen}, \citenamefont {Mao},
  \citenamefont {Huang}, \citenamefont {Zhang},\ and\ \citenamefont
  {et~al.}}]{Yin:16}%
  \BibitemOpen
  \bibfield  {author} {\bibinfo {author} {\bibfnamefont {H.~L.}\ \bibnamefont
  {Yin}}, \bibinfo {author} {\bibfnamefont {T.~Y.}\ \bibnamefont {Chen}},
  \bibinfo {author} {\bibfnamefont {Z.~W.}\ \bibnamefont {Yu}}, \bibinfo
  {author} {\bibfnamefont {H.}~\bibnamefont {Liu}}, \bibinfo {author}
  {\bibfnamefont {L.~X.}\ \bibnamefont {You}}, \bibinfo {author} {\bibfnamefont
  {Y.~H.}\ \bibnamefont {Zhou}}, \bibinfo {author} {\bibfnamefont {S.~J.}\
  \bibnamefont {Chen}}, \bibinfo {author} {\bibfnamefont {Y.~Q.}\ \bibnamefont
  {Mao}}, \bibinfo {author} {\bibfnamefont {M.~Q.}\ \bibnamefont {Huang}},
  \bibinfo {author} {\bibfnamefont {W.~J.}\ \bibnamefont {Zhang}}, \ and\
  \bibinfo {author} {\bibnamefont {et~al.}},\ }\href {\doibase
  10.1103/PhysRevLett.117.190501} {\bibfield  {journal} {\bibinfo  {journal}
  {Phys. Rev. Lett.}\ }\textbf {\bibinfo {volume} {117}},\ \bibinfo {pages}
  {190501} (\bibinfo {year} {2016})}\BibitemShut {NoStop}%
\bibitem [{\citenamefont {Cao}\ \emph {et~al.}(2020)\citenamefont {Cao},
  \citenamefont {Li}, \citenamefont {Yang}, \citenamefont {Jiang},
  \citenamefont {Li}, \citenamefont {Hu}, \citenamefont {Abulizi},
  \citenamefont {Li}, \citenamefont {Zhang}, \citenamefont {Sun},\ and\
  \citenamefont {et~al.}}]{Cao:20}%
  \BibitemOpen
  \bibfield  {author} {\bibinfo {author} {\bibfnamefont {Y.}~\bibnamefont
  {Cao}}, \bibinfo {author} {\bibfnamefont {Y.~H.}\ \bibnamefont {Li}},
  \bibinfo {author} {\bibfnamefont {K.~X.}\ \bibnamefont {Yang}}, \bibinfo
  {author} {\bibfnamefont {Y.~F.}\ \bibnamefont {Jiang}}, \bibinfo {author}
  {\bibfnamefont {S.~L.}\ \bibnamefont {Li}}, \bibinfo {author} {\bibfnamefont
  {X.~L.}\ \bibnamefont {Hu}}, \bibinfo {author} {\bibfnamefont
  {M.}~\bibnamefont {Abulizi}}, \bibinfo {author} {\bibfnamefont {C.~L.}\
  \bibnamefont {Li}}, \bibinfo {author} {\bibfnamefont {W.~J.}\ \bibnamefont
  {Zhang}}, \bibinfo {author} {\bibfnamefont {Q.~C.}\ \bibnamefont {Sun}}, \
  and\ \bibinfo {author} {\bibnamefont {et~al.}},\ }\href {\doibase
  10.1103/PhysRevLett.125.260503} {\bibfield  {journal} {\bibinfo  {journal}
  {Phys. Rev. Lett.}\ }\textbf {\bibinfo {volume} {125}},\ \bibinfo {pages}
  {260503} (\bibinfo {year} {2020})}\BibitemShut {NoStop}%
\bibitem [{\citenamefont {Li}\ \emph {et~al.}(2023{\natexlab{b}})\citenamefont
  {Li}, \citenamefont {Li}, \citenamefont {Hu}, \citenamefont {Jiang},
  \citenamefont {Yu}, \citenamefont {Li}, \citenamefont {Liu}, \citenamefont
  {Liao}, \citenamefont {Ren}, \citenamefont {Li},\ and\ \citenamefont
  {et~al.}}]{Li:23}%
  \BibitemOpen
  \bibfield  {author} {\bibinfo {author} {\bibfnamefont {Y.~H.}\ \bibnamefont
  {Li}}, \bibinfo {author} {\bibfnamefont {S.~L.}\ \bibnamefont {Li}}, \bibinfo
  {author} {\bibfnamefont {X.~L.}\ \bibnamefont {Hu}}, \bibinfo {author}
  {\bibfnamefont {C.}~\bibnamefont {Jiang}}, \bibinfo {author} {\bibfnamefont
  {Z.~W.}\ \bibnamefont {Yu}}, \bibinfo {author} {\bibfnamefont
  {W.}~\bibnamefont {Li}}, \bibinfo {author} {\bibfnamefont {W.~Y.}\
  \bibnamefont {Liu}}, \bibinfo {author} {\bibfnamefont {S.~K.}\ \bibnamefont
  {Liao}}, \bibinfo {author} {\bibfnamefont {J.~G.}\ \bibnamefont {Ren}},
  \bibinfo {author} {\bibfnamefont {H.}~\bibnamefont {Li}}, \ and\ \bibinfo
  {author} {\bibnamefont {et~al.}},\ }\href {\doibase
  10.1103/PhysRevLett.131.100802} {\bibfield  {journal} {\bibinfo  {journal}
  {Phys. Rev. Lett.}\ }\textbf {\bibinfo {volume} {131}},\ \bibinfo {pages}
  {100802} (\bibinfo {year} {2023}{\natexlab{b}})}\BibitemShut {NoStop}%
\bibitem [{\citenamefont {Tian}\ \emph {et~al.}(2022)\citenamefont {Tian},
  \citenamefont {Wang}, \citenamefont {Liu}, \citenamefont {Du}, \citenamefont
  {Liu}, \citenamefont {Lu}, \citenamefont {Wang},\ and\ \citenamefont
  {Li}}]{Tian:22}%
  \BibitemOpen
  \bibfield  {author} {\bibinfo {author} {\bibfnamefont {Y.}~\bibnamefont
  {Tian}}, \bibinfo {author} {\bibfnamefont {P.}~\bibnamefont {Wang}}, \bibinfo
  {author} {\bibfnamefont {J.~Q.}\ \bibnamefont {Liu}}, \bibinfo {author}
  {\bibfnamefont {S.~N.}\ \bibnamefont {Du}}, \bibinfo {author} {\bibfnamefont
  {W.~Y.}\ \bibnamefont {Liu}}, \bibinfo {author} {\bibfnamefont {Z.~G.}\
  \bibnamefont {Lu}}, \bibinfo {author} {\bibfnamefont {X.~Y.}\ \bibnamefont
  {Wang}}, \ and\ \bibinfo {author} {\bibfnamefont {Y.~M.}\ \bibnamefont
  {Li}},\ }\href {\doibase 10.1364/OPTICA.450573} {\bibfield  {journal}
  {\bibinfo  {journal} {Optica}\ }\textbf {\bibinfo {volume} {9}},\ \bibinfo
  {pages} {492} (\bibinfo {year} {2022})}\BibitemShut {NoStop}%
\bibitem [{\citenamefont {Pirandola}\ \emph {et~al.}(2015)\citenamefont
  {Pirandola}, \citenamefont {Ottaviani}, \citenamefont {Spedalieri},
  \citenamefont {Weedbrook}, \citenamefont {Braunstein}, \citenamefont {Lloyd},
  \citenamefont {Gehring}, \citenamefont {Jacobsen},\ and\ \citenamefont
  {Andersen}}]{Pirandola:15}%
  \BibitemOpen
  \bibfield  {author} {\bibinfo {author} {\bibfnamefont {S.}~\bibnamefont
  {Pirandola}}, \bibinfo {author} {\bibfnamefont {C.}~\bibnamefont
  {Ottaviani}}, \bibinfo {author} {\bibfnamefont {G.}~\bibnamefont
  {Spedalieri}}, \bibinfo {author} {\bibfnamefont {C.}~\bibnamefont
  {Weedbrook}}, \bibinfo {author} {\bibfnamefont {S.~L.}\ \bibnamefont
  {Braunstein}}, \bibinfo {author} {\bibfnamefont {S.}~\bibnamefont {Lloyd}},
  \bibinfo {author} {\bibfnamefont {T.}~\bibnamefont {Gehring}}, \bibinfo
  {author} {\bibfnamefont {C.~S.}\ \bibnamefont {Jacobsen}}, \ and\ \bibinfo
  {author} {\bibfnamefont {U.~L.}\ \bibnamefont {Andersen}},\ }\href {\doibase
  10.1038/NPHOTON.2015.83} {\bibfield  {journal} {\bibinfo  {journal} {Nat.
  Photonics}\ }\textbf {\bibinfo {volume} {9}},\ \bibinfo {pages} {397}
  (\bibinfo {year} {2015})}\BibitemShut {NoStop}%
\bibitem [{\citenamefont {Gisin}\ \emph {et~al.}(2006)\citenamefont {Gisin},
  \citenamefont {Fasel}, \citenamefont {Kraus}, \citenamefont {Zbinden},\ and\
  \citenamefont {Ribordy}}]{Gisin:06}%
  \BibitemOpen
  \bibfield  {author} {\bibinfo {author} {\bibfnamefont {N.}~\bibnamefont
  {Gisin}}, \bibinfo {author} {\bibfnamefont {S.}~\bibnamefont {Fasel}},
  \bibinfo {author} {\bibfnamefont {B.}~\bibnamefont {Kraus}}, \bibinfo
  {author} {\bibfnamefont {H.}~\bibnamefont {Zbinden}}, \ and\ \bibinfo
  {author} {\bibfnamefont {G.}~\bibnamefont {Ribordy}},\ }\href {\doibase
  10.1103/PhysRevA.73.022320} {\bibfield  {journal} {\bibinfo  {journal} {Phys.
  Rev. A}\ }\textbf {\bibinfo {volume} {73}},\ \bibinfo {pages} {022320}
  (\bibinfo {year} {2006})}\BibitemShut {NoStop}%
\bibitem [{\citenamefont {Khan}\ \emph {et~al.}(2014)\citenamefont {Khan},
  \citenamefont {Stiller}, \citenamefont {Jain}, \citenamefont {Jouguet},
  \citenamefont {Kunz-Jacques}, \citenamefont {Diamanti}, \citenamefont
  {Marquardt},\ and\ \citenamefont {Leuchs}}]{Khan:14}%
  \BibitemOpen
  \bibfield  {author} {\bibinfo {author} {\bibfnamefont {I.}~\bibnamefont
  {Khan}}, \bibinfo {author} {\bibfnamefont {B.}~\bibnamefont {Stiller}},
  \bibinfo {author} {\bibfnamefont {N.}~\bibnamefont {Jain}}, \bibinfo {author}
  {\bibfnamefont {P.}~\bibnamefont {Jouguet}}, \bibinfo {author} {\bibfnamefont
  {S.}~\bibnamefont {Kunz-Jacques}}, \bibinfo {author} {\bibfnamefont
  {E.}~\bibnamefont {Diamanti}}, \bibinfo {author} {\bibfnamefont
  {C.}~\bibnamefont {Marquardt}}, \ and\ \bibinfo {author} {\bibfnamefont
  {G.}~\bibnamefont {Leuchs}},\ }in\ \href {https://qcrypt.net/2014/} {\emph
  {\bibinfo {booktitle} {Conference on Quantum Cryptography (QCRYPT)}}}\
  (\bibinfo {address} {Paris, France},\ \bibinfo {year} {2014})\BibitemShut
  {NoStop}%
\bibitem [{\citenamefont {Jouguet}\ \emph {et~al.}(2012)\citenamefont
  {Jouguet}, \citenamefont {Kunz-Jacques}, \citenamefont {Diamanti},\ and\
  \citenamefont {Leverrier}}]{Jouguet:12}%
  \BibitemOpen
  \bibfield  {author} {\bibinfo {author} {\bibfnamefont {P.}~\bibnamefont
  {Jouguet}}, \bibinfo {author} {\bibfnamefont {S.}~\bibnamefont
  {Kunz-Jacques}}, \bibinfo {author} {\bibfnamefont {E.}~\bibnamefont
  {Diamanti}}, \ and\ \bibinfo {author} {\bibfnamefont {A.}~\bibnamefont
  {Leverrier}},\ }\href {\doibase 10.1103/PhysRevA.86.032309} {\bibfield
  {journal} {\bibinfo  {journal} {Phys. Rev. A}\ }\textbf {\bibinfo {volume}
  {86}},\ \bibinfo {pages} {032309} (\bibinfo {year} {2012})}\BibitemShut
  {NoStop}%
\bibitem [{\citenamefont {Liu}\ \emph {et~al.}(2017)\citenamefont {Liu},
  \citenamefont {Wang}, \citenamefont {Wang}, \citenamefont {Du},\ and\
  \citenamefont {Li}}]{Liu:17}%
  \BibitemOpen
  \bibfield  {author} {\bibinfo {author} {\bibfnamefont {W.~Y.}\ \bibnamefont
  {Liu}}, \bibinfo {author} {\bibfnamefont {X.~Y.}\ \bibnamefont {Wang}},
  \bibinfo {author} {\bibfnamefont {N.}~\bibnamefont {Wang}}, \bibinfo {author}
  {\bibfnamefont {S.~N.}\ \bibnamefont {Du}}, \ and\ \bibinfo {author}
  {\bibfnamefont {Y.~M.}\ \bibnamefont {Li}},\ }\href {\doibase
  10.1103/PhysRevA.96.042312} {\bibfield  {journal} {\bibinfo  {journal} {Phys.
  Rev. A}\ }\textbf {\bibinfo {volume} {96}},\ \bibinfo {pages} {042312}
  (\bibinfo {year} {2017})}\BibitemShut {NoStop}%
\bibitem [{\citenamefont {Yoshino}\ \emph {et~al.}(2018)\citenamefont
  {Yoshino}, \citenamefont {Fujiwara}, \citenamefont {Nakata}, \citenamefont
  {Sumiya}, \citenamefont {Sasaki}, \citenamefont {Takeoka}, \citenamefont
  {Sasaki}, \citenamefont {Tajima}, \citenamefont {Koashi},\ and\ \citenamefont
  {Tomita}}]{Yoshino:18}%
  \BibitemOpen
  \bibfield  {author} {\bibinfo {author} {\bibfnamefont {K.~I.}\ \bibnamefont
  {Yoshino}}, \bibinfo {author} {\bibfnamefont {M.}~\bibnamefont {Fujiwara}},
  \bibinfo {author} {\bibfnamefont {K.}~\bibnamefont {Nakata}}, \bibinfo
  {author} {\bibfnamefont {T.}~\bibnamefont {Sumiya}}, \bibinfo {author}
  {\bibfnamefont {T.}~\bibnamefont {Sasaki}}, \bibinfo {author} {\bibfnamefont
  {M.}~\bibnamefont {Takeoka}}, \bibinfo {author} {\bibfnamefont
  {M.}~\bibnamefont {Sasaki}}, \bibinfo {author} {\bibfnamefont
  {A.}~\bibnamefont {Tajima}}, \bibinfo {author} {\bibfnamefont
  {M.}~\bibnamefont {Koashi}}, \ and\ \bibinfo {author} {\bibfnamefont
  {A.}~\bibnamefont {Tomita}},\ }\href {\doibase 10.1038/s41534-017-0057-8}
  {\bibfield  {journal} {\bibinfo  {journal} {npj Quantum Inf.}\ }\textbf
  {\bibinfo {volume} {4}} (\bibinfo {year} {2018}),\
  10.1038/s41534-017-0057-8}\BibitemShut {NoStop}%
\bibitem [{\citenamefont {Laudenbach}\ \emph
  {et~al.}(2018{\natexlab{a}})\citenamefont {Laudenbach}, \citenamefont
  {Pacher}, \citenamefont {Fung}, \citenamefont {Poppe}, \citenamefont {Peev},
  \citenamefont {Schrenk}, \citenamefont {Hentschel}, \citenamefont {Walther},\
  and\ \citenamefont {H\"ubel}}]{Laudenbach:18}%
  \BibitemOpen
  \bibfield  {author} {\bibinfo {author} {\bibfnamefont {F.}~\bibnamefont
  {Laudenbach}}, \bibinfo {author} {\bibfnamefont {C.}~\bibnamefont {Pacher}},
  \bibinfo {author} {\bibfnamefont {C.-H.~F.}\ \bibnamefont {Fung}}, \bibinfo
  {author} {\bibfnamefont {A.}~\bibnamefont {Poppe}}, \bibinfo {author}
  {\bibfnamefont {M.}~\bibnamefont {Peev}}, \bibinfo {author} {\bibfnamefont
  {B.}~\bibnamefont {Schrenk}}, \bibinfo {author} {\bibfnamefont
  {M.}~\bibnamefont {Hentschel}}, \bibinfo {author} {\bibfnamefont
  {P.}~\bibnamefont {Walther}}, \ and\ \bibinfo {author} {\bibfnamefont
  {H.}~\bibnamefont {H\"ubel}},\ }\href {\doibase 10.1002/qute.201800011}
  {\bibfield  {journal} {\bibinfo  {journal} {Adv. Quantum Technol.}\ }\textbf
  {\bibinfo {volume} {1}},\ \bibinfo {pages} {1800011} (\bibinfo {year}
  {2018}{\natexlab{a}})}\BibitemShut {NoStop}%
\bibitem [{\citenamefont {Li}\ \emph {et~al.}(2021)\citenamefont {Li},
  \citenamefont {Qian},\ and\ \citenamefont {Lo}}]{Li:21}%
  \BibitemOpen
  \bibfield  {author} {\bibinfo {author} {\bibfnamefont {C.~Y.}\ \bibnamefont
  {Li}}, \bibinfo {author} {\bibfnamefont {L.}~\bibnamefont {Qian}}, \ and\
  \bibinfo {author} {\bibfnamefont {H.-K.}\ \bibnamefont {Lo}},\ }\href
  {\doibase 10.1038/s41534-021-00482-3} {\bibfield  {journal} {\bibinfo
  {journal} {npj Quantum Inf.}\ }\textbf {\bibinfo {volume} {7}} (\bibinfo
  {year} {2021}),\ 10.1038/s41534-021-00482-3}\BibitemShut {NoStop}%
\bibitem [{\citenamefont {Huang}\ \emph {et~al.}(2016)\citenamefont {Huang},
  \citenamefont {Huang}, \citenamefont {k.~Lin},\ and\ \citenamefont
  {Zeng}}]{Huang:16}%
  \BibitemOpen
  \bibfield  {author} {\bibinfo {author} {\bibfnamefont {D.}~\bibnamefont
  {Huang}}, \bibinfo {author} {\bibfnamefont {P.}~\bibnamefont {Huang}},
  \bibinfo {author} {\bibfnamefont {D.}~\bibnamefont {k.~Lin}}, \ and\ \bibinfo
  {author} {\bibfnamefont {G.~H.}\ \bibnamefont {Zeng}},\ }\href {\doibase
  10.1038/srep19201} {\bibfield  {journal} {\bibinfo  {journal} {Sci. Rep.}\
  }\textbf {\bibinfo {volume} {6}} (\bibinfo {year} {2016}),\
  10.1038/srep19201}\BibitemShut {NoStop}%
\bibitem [{\citenamefont {Derkach}\ \emph {et~al.}(2017)\citenamefont
  {Derkach}, \citenamefont {Usenko},\ and\ \citenamefont {Filip}}]{Derkach:17}%
  \BibitemOpen
  \bibfield  {author} {\bibinfo {author} {\bibfnamefont {I.}~\bibnamefont
  {Derkach}}, \bibinfo {author} {\bibfnamefont {V.~C.}\ \bibnamefont {Usenko}},
  \ and\ \bibinfo {author} {\bibfnamefont {R.}~\bibnamefont {Filip}},\ }\href
  {\doibase 10.1103/PhysRevA.96.062309} {\bibfield  {journal} {\bibinfo
  {journal} {Phys. Rev. A}\ }\textbf {\bibinfo {volume} {96}},\ \bibinfo
  {pages} {062309} (\bibinfo {year} {2017})}\BibitemShut {NoStop}%
\bibitem [{\citenamefont {Usenko}\ and\ \citenamefont
  {Filip}(2010)}]{Usenko:10}%
  \BibitemOpen
  \bibfield  {author} {\bibinfo {author} {\bibfnamefont {V.~C.}\ \bibnamefont
  {Usenko}}\ and\ \bibinfo {author} {\bibfnamefont {R.}~\bibnamefont {Filip}},\
  }\href {\doibase 10.1103/PhysRevA.81.022318} {\bibfield  {journal} {\bibinfo
  {journal} {Phys. Rev. A}\ }\textbf {\bibinfo {volume} {81}},\ \bibinfo
  {pages} {022318} (\bibinfo {year} {2010})}\BibitemShut {NoStop}%
\bibitem [{\citenamefont {Derkach}\ \emph {et~al.}(2016)\citenamefont
  {Derkach}, \citenamefont {Usenko},\ and\ \citenamefont {Filip}}]{Derkach:16}%
  \BibitemOpen
  \bibfield  {author} {\bibinfo {author} {\bibfnamefont {I.}~\bibnamefont
  {Derkach}}, \bibinfo {author} {\bibfnamefont {V.~C.}\ \bibnamefont {Usenko}},
  \ and\ \bibinfo {author} {\bibfnamefont {R.}~\bibnamefont {Filip}},\ }\href
  {\doibase 10.1103/PhysRevA.93.032309} {\bibfield  {journal} {\bibinfo
  {journal} {Phys. Rev. A}\ }\textbf {\bibinfo {volume} {93}},\ \bibinfo
  {pages} {032309} (\bibinfo {year} {2016})}\BibitemShut {NoStop}%
\bibitem [{\citenamefont {Tamaki}\ \emph {et~al.}(2016)\citenamefont {Tamaki},
  \citenamefont {Curty},\ and\ \citenamefont {Lucamarini}}]{Tamaki:16}%
  \BibitemOpen
  \bibfield  {author} {\bibinfo {author} {\bibfnamefont {K.}~\bibnamefont
  {Tamaki}}, \bibinfo {author} {\bibfnamefont {M.}~\bibnamefont {Curty}}, \
  and\ \bibinfo {author} {\bibfnamefont {M.}~\bibnamefont {Lucamarini}},\
  }\href {\doibase 10.1088/1367-2630/18/6/065008} {\bibfield  {journal}
  {\bibinfo  {journal} {New J. Phys.}\ }\textbf {\bibinfo {volume} {18}},\
  \bibinfo {pages} {065008} (\bibinfo {year} {2016})}\BibitemShut {NoStop}%
\bibitem [{\citenamefont {Pereira}\ \emph {et~al.}(2019)\citenamefont
  {Pereira}, \citenamefont {Curty},\ and\ \citenamefont {Tamaki}}]{Pereira:19}%
  \BibitemOpen
  \bibfield  {author} {\bibinfo {author} {\bibfnamefont {M.}~\bibnamefont
  {Pereira}}, \bibinfo {author} {\bibfnamefont {M.}~\bibnamefont {Curty}}, \
  and\ \bibinfo {author} {\bibfnamefont {K.}~\bibnamefont {Tamaki}},\ }\href
  {\doibase 10.1038/s41534-019-0180-9} {\bibfield  {journal} {\bibinfo
  {journal} {npj Quantum Inf.}\ }\textbf {\bibinfo {volume} {5}} (\bibinfo
  {year} {2019}),\ 10.1038/s41534-019-0180-9}\BibitemShut {NoStop}%
\bibitem [{\citenamefont {Hajomer}\ \emph {et~al.}(2022)\citenamefont
  {Hajomer}, \citenamefont {Jain}, \citenamefont {Mani}, \citenamefont {Chin},
  \citenamefont {Andersen},\ and\ \citenamefont {Gehring}}]{Hajomer:22}%
  \BibitemOpen
  \bibfield  {author} {\bibinfo {author} {\bibfnamefont {A.~A.~E.}\
  \bibnamefont {Hajomer}}, \bibinfo {author} {\bibfnamefont {N.}~\bibnamefont
  {Jain}}, \bibinfo {author} {\bibfnamefont {H.}~\bibnamefont {Mani}}, \bibinfo
  {author} {\bibfnamefont {H.-M.}\ \bibnamefont {Chin}}, \bibinfo {author}
  {\bibfnamefont {U.~L.}\ \bibnamefont {Andersen}}, \ and\ \bibinfo {author}
  {\bibfnamefont {T.}~\bibnamefont {Gehring}},\ }\href {\doibase
  10.1038/s41534-022-00640-1} {\bibfield  {journal} {\bibinfo  {journal} {npj
  Quantum Inf.}\ }\textbf {\bibinfo {volume} {8}},\ \bibinfo {pages} {136}
  (\bibinfo {year} {2022})}\BibitemShut {NoStop}%
\bibitem [{\citenamefont {Wang}\ \emph
  {et~al.}(2023{\natexlab{b}})\citenamefont {Wang}, \citenamefont {Wang},
  \citenamefont {Hu}, \citenamefont {Zapatero}, \citenamefont {Qian},
  \citenamefont {Qi},\ and\ \citenamefont {Lo}}]{Wang:2023}%
  \BibitemOpen
  \bibfield  {author} {\bibinfo {author} {\bibfnamefont {W.}~\bibnamefont
  {Wang}}, \bibinfo {author} {\bibfnamefont {R.}~\bibnamefont {Wang}}, \bibinfo
  {author} {\bibfnamefont {C.}~\bibnamefont {Hu}}, \bibinfo {author}
  {\bibfnamefont {V.}~\bibnamefont {Zapatero}}, \bibinfo {author}
  {\bibfnamefont {L.}~\bibnamefont {Qian}}, \bibinfo {author} {\bibfnamefont
  {B.}~\bibnamefont {Qi}}, \ and\ \bibinfo {author} {\bibfnamefont {H.-K.}\
  \bibnamefont {Lo}},\ }\href {\doibase 10.1103/PhysRevLett.130.220801}
  {\bibfield  {journal} {\bibinfo  {journal} {Phys. Rev. Lett.}\ }\textbf
  {\bibinfo {volume} {130}},\ \bibinfo {pages} {220801} (\bibinfo {year}
  {2023}{\natexlab{b}})}\BibitemShut {NoStop}%
\bibitem [{\citenamefont {Lu}\ \emph {et~al.}(2023)\citenamefont {Lu},
  \citenamefont {Wang}, \citenamefont {Zapatero}, \citenamefont {Chen},
  \citenamefont {Wang}, \citenamefont {Yin}, \citenamefont {Curty},
  \citenamefont {He}, \citenamefont {Wang}, \citenamefont {Chen},\ and\
  \citenamefont {et~al.}}]{Lu:23}%
  \BibitemOpen
  \bibfield  {author} {\bibinfo {author} {\bibfnamefont {F.~Y.}\ \bibnamefont
  {Lu}}, \bibinfo {author} {\bibfnamefont {Z.~H.}\ \bibnamefont {Wang}},
  \bibinfo {author} {\bibfnamefont {V.}~\bibnamefont {Zapatero}}, \bibinfo
  {author} {\bibfnamefont {J.~L.}\ \bibnamefont {Chen}}, \bibinfo {author}
  {\bibfnamefont {S.}~\bibnamefont {Wang}}, \bibinfo {author} {\bibfnamefont
  {Z.~Q.}\ \bibnamefont {Yin}}, \bibinfo {author} {\bibfnamefont
  {M.}~\bibnamefont {Curty}}, \bibinfo {author} {\bibfnamefont {D.~Y.}\
  \bibnamefont {He}}, \bibinfo {author} {\bibfnamefont {R.}~\bibnamefont
  {Wang}}, \bibinfo {author} {\bibfnamefont {W.}~\bibnamefont {Chen}}, \ and\
  \bibinfo {author} {\bibnamefont {et~al.}},\ }\href {\doibase
  10.1103/PhysRevLett.131.110802} {\bibfield  {journal} {\bibinfo  {journal}
  {Phys. Rev. Lett.}\ }\textbf {\bibinfo {volume} {131}},\ \bibinfo {pages}
  {110802} (\bibinfo {year} {2023})}\BibitemShut {NoStop}%
\bibitem [{\citenamefont {Hu}\ \emph {et~al.}(2023)\citenamefont {Hu},
  \citenamefont {Wang}, \citenamefont {Chan}, \citenamefont {Yuan},\ and\
  \citenamefont {Lo}}]{Hu:23}%
  \BibitemOpen
  \bibfield  {author} {\bibinfo {author} {\bibfnamefont {C.~Q.}\ \bibnamefont
  {Hu}}, \bibinfo {author} {\bibfnamefont {W.~Y.}\ \bibnamefont {Wang}},
  \bibinfo {author} {\bibfnamefont {K.-S.}\ \bibnamefont {Chan}}, \bibinfo
  {author} {\bibfnamefont {Z.~H.}\ \bibnamefont {Yuan}}, \ and\ \bibinfo
  {author} {\bibfnamefont {H.-K.}\ \bibnamefont {Lo}},\ }\href {\doibase
  10.1103/PhysRevLett.131.110801} {\bibfield  {journal} {\bibinfo  {journal}
  {Phys. Rev. Lett.}\ }\textbf {\bibinfo {volume} {131}},\ \bibinfo {pages}
  {110801} (\bibinfo {year} {2023})}\BibitemShut {NoStop}%
\bibitem [{\citenamefont {Zapatero}\ \emph {et~al.}(2023)\citenamefont
  {Zapatero}, \citenamefont {Wang},\ and\ \citenamefont {Curty}}]{Zapatero:23}%
  \BibitemOpen
  \bibfield  {author} {\bibinfo {author} {\bibfnamefont {V.}~\bibnamefont
  {Zapatero}}, \bibinfo {author} {\bibfnamefont {W.~Y.}\ \bibnamefont {Wang}},
  \ and\ \bibinfo {author} {\bibfnamefont {M.}~\bibnamefont {Curty}},\ }\href
  {\doibase 10.1088/2058-9565/acbc46} {\bibfield  {journal} {\bibinfo
  {journal} {Quantum Sci. Technol.}\ }\textbf {\bibinfo {volume} {8}},\
  \bibinfo {pages} {025014} (\bibinfo {year} {2023})}\BibitemShut {NoStop}%
\bibitem [{\citenamefont {Qi}\ \emph {et~al.}(2020)\citenamefont {Qi},
  \citenamefont {Gunther}, \citenamefont {Evans}, \citenamefont {Williams},
  \citenamefont {Camacho},\ and\ \citenamefont {Peters}}]{Qi:20}%
  \BibitemOpen
  \bibfield  {author} {\bibinfo {author} {\bibfnamefont {B.}~\bibnamefont
  {Qi}}, \bibinfo {author} {\bibfnamefont {H.}~\bibnamefont {Gunther}},
  \bibinfo {author} {\bibfnamefont {P.~G.}\ \bibnamefont {Evans}}, \bibinfo
  {author} {\bibfnamefont {B.~P.}\ \bibnamefont {Williams}}, \bibinfo {author}
  {\bibfnamefont {R.~M.}\ \bibnamefont {Camacho}}, \ and\ \bibinfo {author}
  {\bibfnamefont {N.~A.}\ \bibnamefont {Peters}},\ }\href {\doibase
  10.1103/PhysRevApplied.13.054065} {\bibfield  {journal} {\bibinfo  {journal}
  {Phys. Rev. Appl.}\ }\textbf {\bibinfo {volume} {13}},\ \bibinfo {pages}
  {054065} (\bibinfo {year} {2020})}\BibitemShut {NoStop}%
\bibitem [{\citenamefont {Qi}\ \emph {et~al.}(2018)\citenamefont {Qi},
  \citenamefont {Evans},\ and\ \citenamefont {Grice}}]{Qi:18}%
  \BibitemOpen
  \bibfield  {author} {\bibinfo {author} {\bibfnamefont {B.}~\bibnamefont
  {Qi}}, \bibinfo {author} {\bibfnamefont {P.~G.}\ \bibnamefont {Evans}}, \
  and\ \bibinfo {author} {\bibfnamefont {W.~P.}\ \bibnamefont {Grice}},\ }\href
  {\doibase 10.1103/PhysRevA.97.012317} {\bibfield  {journal} {\bibinfo
  {journal} {Phys. Rev. A}\ }\textbf {\bibinfo {volume} {97}},\ \bibinfo
  {pages} {012317} (\bibinfo {year} {2018})}\BibitemShut {NoStop}%
\bibitem [{\citenamefont {Huang}\ \emph
  {et~al.}(2021{\natexlab{b}})\citenamefont {Huang}, \citenamefont {Wang},
  \citenamefont {Chen}, \citenamefont {Wang}, \citenamefont {Zhou},\ and\
  \citenamefont {Zeng}}]{Huang:21}%
  \BibitemOpen
  \bibfield  {author} {\bibinfo {author} {\bibfnamefont {P.}~\bibnamefont
  {Huang}}, \bibinfo {author} {\bibfnamefont {T.}~\bibnamefont {Wang}},
  \bibinfo {author} {\bibfnamefont {R.}~\bibnamefont {Chen}}, \bibinfo {author}
  {\bibfnamefont {P.}~\bibnamefont {Wang}}, \bibinfo {author} {\bibfnamefont
  {Y.~M.}\ \bibnamefont {Zhou}}, \ and\ \bibinfo {author} {\bibfnamefont
  {G.~H.}\ \bibnamefont {Zeng}},\ }\href {\doibase 10.1088/1367-2630/ac3684}
  {\bibfield  {journal} {\bibinfo  {journal} {New J. Phys.}\ }\textbf {\bibinfo
  {volume} {23}},\ \bibinfo {pages} {113028} (\bibinfo {year}
  {2021}{\natexlab{b}})}\BibitemShut {NoStop}%
\bibitem [{\citenamefont {Wu}\ \emph {et~al.}(2021)\citenamefont {Wu},
  \citenamefont {J.Wang}, \citenamefont {Guo}, \citenamefont {Zhong},\ and\
  \citenamefont {Huang}}]{Wu:21}%
  \BibitemOpen
  \bibfield  {author} {\bibinfo {author} {\bibfnamefont {X.~D.}\ \bibnamefont
  {Wu}}, \bibinfo {author} {\bibfnamefont {Y.}~\bibnamefont {J.Wang}}, \bibinfo
  {author} {\bibfnamefont {Y.}~\bibnamefont {Guo}}, \bibinfo {author}
  {\bibfnamefont {H.}~\bibnamefont {Zhong}}, \ and\ \bibinfo {author}
  {\bibfnamefont {D.}~\bibnamefont {Huang}},\ }\href {\doibase
  10.1103/PhysRevA.103.032604} {\bibfield  {journal} {\bibinfo  {journal}
  {Phys. Rev. A}\ }\textbf {\bibinfo {volume} {103}},\ \bibinfo {pages}
  {032604} (\bibinfo {year} {2021})}\BibitemShut {NoStop}%
\bibitem [{\citenamefont {Li}\ \emph {et~al.}(2022)\citenamefont {Li},
  \citenamefont {Hu}, \citenamefont {Wang}, \citenamefont {Wang},\ and\
  \citenamefont {Lo}}]{Li2022Passive}%
  \BibitemOpen
  \bibfield  {author} {\bibinfo {author} {\bibfnamefont {C.}~\bibnamefont
  {Li}}, \bibinfo {author} {\bibfnamefont {C.}~\bibnamefont {Hu}}, \bibinfo
  {author} {\bibfnamefont {W.}~\bibnamefont {Wang}}, \bibinfo {author}
  {\bibfnamefont {R.}~\bibnamefont {Wang}}, \ and\ \bibinfo {author}
  {\bibfnamefont {H.-K.}\ \bibnamefont {Lo}},\ }\href {\doibase
  10.48550/arXiv.2212.01876} {\bibfield  {journal} {\bibinfo  {journal} {arXiv
  preprint}\ ,\ \bibinfo {pages} {5}} (\bibinfo {year} {2022})},\ \Eprint
  {http://arxiv.org/abs/2212.01876} {arXiv:2212.01876 [quant-ph]} \BibitemShut
  {NoStop}%
\bibitem [{\citenamefont {Tian}\ \emph {et~al.}(2023)\citenamefont {Tian},
  \citenamefont {Zhang}, \citenamefont {Liu}, \citenamefont {Wang},
  \citenamefont {Lu}, \citenamefont {Wang},\ and\ \citenamefont
  {Li}}]{Tian:23}%
  \BibitemOpen
  \bibfield  {author} {\bibinfo {author} {\bibfnamefont {Y.}~\bibnamefont
  {Tian}}, \bibinfo {author} {\bibfnamefont {Y.}~\bibnamefont {Zhang}},
  \bibinfo {author} {\bibfnamefont {S.~S.}\ \bibnamefont {Liu}}, \bibinfo
  {author} {\bibfnamefont {P.}~\bibnamefont {Wang}}, \bibinfo {author}
  {\bibfnamefont {Z.~G.}\ \bibnamefont {Lu}}, \bibinfo {author} {\bibfnamefont
  {X.~Y.}\ \bibnamefont {Wang}}, \ and\ \bibinfo {author} {\bibfnamefont
  {Y.~M.}\ \bibnamefont {Li}},\ }\href {\doibase 10.1364/OL.492082} {\bibfield
  {journal} {\bibinfo  {journal} {Opt. Lett.}\ }\textbf {\bibinfo {volume}
  {48}},\ \bibinfo {pages} {2953} (\bibinfo {year} {2023})}\BibitemShut
  {NoStop}%
\bibitem [{\citenamefont {Wang}\ \emph {et~al.}(2022)\citenamefont {Wang},
  \citenamefont {Li}, \citenamefont {Pi}, \citenamefont {Pan}, \citenamefont
  {Shao}, \citenamefont {Ma}, \citenamefont {Zhang}, \citenamefont {Yang},
  \citenamefont {Zhang}, \citenamefont {Huang},\ and\ \citenamefont
  {et~al.}}]{Wang:22}%
  \BibitemOpen
  \bibfield  {author} {\bibinfo {author} {\bibfnamefont {H.}~\bibnamefont
  {Wang}}, \bibinfo {author} {\bibfnamefont {Y.}~\bibnamefont {Li}}, \bibinfo
  {author} {\bibfnamefont {Y.~D.}\ \bibnamefont {Pi}}, \bibinfo {author}
  {\bibfnamefont {Y.}~\bibnamefont {Pan}}, \bibinfo {author} {\bibfnamefont
  {Y.}~\bibnamefont {Shao}}, \bibinfo {author} {\bibfnamefont {L.}~\bibnamefont
  {Ma}}, \bibinfo {author} {\bibfnamefont {Y.~C.}\ \bibnamefont {Zhang}},
  \bibinfo {author} {\bibfnamefont {J.}~\bibnamefont {Yang}}, \bibinfo {author}
  {\bibfnamefont {T.}~\bibnamefont {Zhang}}, \bibinfo {author} {\bibfnamefont
  {W.}~\bibnamefont {Huang}}, \ and\ \bibinfo {author} {\bibnamefont
  {et~al.}},\ }\href {\doibase 10.1038/s42005-022-00941-z} {\bibfield
  {journal} {\bibinfo  {journal} {Commun. Phys.}\ }\textbf {\bibinfo {volume}
  {5}},\ \bibinfo {pages} {162} (\bibinfo {year} {2022})}\BibitemShut {NoStop}%
\bibitem [{\citenamefont {Pan}\ \emph {et~al.}(2022)\citenamefont {Pan},
  \citenamefont {Wang}, \citenamefont {Shao}, \citenamefont {Pi}, \citenamefont
  {Li}, \citenamefont {Liu}, \citenamefont {Huang},\ and\ \citenamefont
  {Xu}}]{Pan:22}%
  \BibitemOpen
  \bibfield  {author} {\bibinfo {author} {\bibfnamefont {Y.}~\bibnamefont
  {Pan}}, \bibinfo {author} {\bibfnamefont {H.}~\bibnamefont {Wang}}, \bibinfo
  {author} {\bibfnamefont {Y.}~\bibnamefont {Shao}}, \bibinfo {author}
  {\bibfnamefont {Y.~D.}\ \bibnamefont {Pi}}, \bibinfo {author} {\bibfnamefont
  {Y.}~\bibnamefont {Li}}, \bibinfo {author} {\bibfnamefont {B.}~\bibnamefont
  {Liu}}, \bibinfo {author} {\bibfnamefont {W.}~\bibnamefont {Huang}}, \ and\
  \bibinfo {author} {\bibfnamefont {B.~J.}\ \bibnamefont {Xu}},\ }\href
  {\doibase 10.1364/OL.456978} {\bibfield  {journal} {\bibinfo  {journal} {Opt.
  Lett.}\ }\textbf {\bibinfo {volume} {47}},\ \bibinfo {pages} {3307} (\bibinfo
  {year} {2022})}\BibitemShut {NoStop}%
\bibitem [{\citenamefont {Comandar}\ \emph {et~al.}(2016)\citenamefont
  {Comandar}, \citenamefont {Lucamarini}, \citenamefont {Fröhlich},
  \citenamefont {Dynes}, \citenamefont {Yuan},\ and\ \citenamefont
  {Shields}}]{Comandar:16}%
  \BibitemOpen
  \bibfield  {author} {\bibinfo {author} {\bibfnamefont {L.~C.}\ \bibnamefont
  {Comandar}}, \bibinfo {author} {\bibfnamefont {M.}~\bibnamefont
  {Lucamarini}}, \bibinfo {author} {\bibfnamefont {B.}~\bibnamefont
  {Fröhlich}}, \bibinfo {author} {\bibfnamefont {J.~F.}\ \bibnamefont
  {Dynes}}, \bibinfo {author} {\bibfnamefont {Z.~L.}\ \bibnamefont {Yuan}}, \
  and\ \bibinfo {author} {\bibfnamefont {A.~J.}\ \bibnamefont {Shields}},\
  }\href {https://arxiv.org/abs/1605.04759} {\bibfield  {journal} {\bibinfo
  {journal} {Opt. Express}\ }\textbf {\bibinfo {volume} {24}},\ \bibinfo
  {pages} {17849} (\bibinfo {year} {2016})}\BibitemShut {NoStop}%
\bibitem [{\citenamefont {Yuan}\ \emph
  {et~al.}(2014{\natexlab{a}})\citenamefont {Yuan}, \citenamefont {Lucamarini},
  \citenamefont {Dynes}, \citenamefont {Fröhlich}, \citenamefont {Ward},\ and\
  \citenamefont {Shields}}]{Yuan:14}%
  \BibitemOpen
  \bibfield  {author} {\bibinfo {author} {\bibfnamefont {Z.~L.}\ \bibnamefont
  {Yuan}}, \bibinfo {author} {\bibfnamefont {M.}~\bibnamefont {Lucamarini}},
  \bibinfo {author} {\bibfnamefont {J.~F.}\ \bibnamefont {Dynes}}, \bibinfo
  {author} {\bibfnamefont {B.}~\bibnamefont {Fröhlich}}, \bibinfo {author}
  {\bibfnamefont {M.~B.}\ \bibnamefont {Ward}}, \ and\ \bibinfo {author}
  {\bibfnamefont {A.~J.}\ \bibnamefont {Shields}},\ }\href {\doibase
  10.1103/PhysRevApplied.2.064006} {\bibfield  {journal} {\bibinfo  {journal}
  {Phys. Rev. Appl.}\ }\textbf {\bibinfo {volume} {2}},\ \bibinfo {pages}
  {064006} (\bibinfo {year} {2014}{\natexlab{a}})}\BibitemShut {NoStop}%
\bibitem [{\citenamefont {Yuan}\ \emph
  {et~al.}(2014{\natexlab{b}})\citenamefont {Yuan}, \citenamefont {Lucamarini},
  \citenamefont {Dynes}, \citenamefont {Fröhlich}, \citenamefont {Plews},\
  and\ \citenamefont {Shields}}]{Yuan:2014}%
  \BibitemOpen
  \bibfield  {author} {\bibinfo {author} {\bibfnamefont {Z.~L.}\ \bibnamefont
  {Yuan}}, \bibinfo {author} {\bibfnamefont {M.}~\bibnamefont {Lucamarini}},
  \bibinfo {author} {\bibfnamefont {J.~F.}\ \bibnamefont {Dynes}}, \bibinfo
  {author} {\bibfnamefont {B.}~\bibnamefont {Fröhlich}}, \bibinfo {author}
  {\bibfnamefont {A.}~\bibnamefont {Plews}}, \ and\ \bibinfo {author}
  {\bibfnamefont {A.~J.}\ \bibnamefont {Shields}},\ }\href {\doibase
  10.1063/1.4885874} {\bibfield  {journal} {\bibinfo  {journal} {Appl. Phys.
  Lett.}\ }\textbf {\bibinfo {volume} {104}},\ \bibinfo {pages} {261112}
  (\bibinfo {year} {2014}{\natexlab{b}})}\BibitemShut {NoStop}%
\bibitem [{\citenamefont {Yuan}\ \emph {et~al.}(2016)\citenamefont {Yuan},
  \citenamefont {Fröhlich}, \citenamefont {Lucamarini}, \citenamefont
  {Roberts}, \citenamefont {Dynes},\ and\ \citenamefont {Shields}}]{Yuan:16}%
  \BibitemOpen
  \bibfield  {author} {\bibinfo {author} {\bibfnamefont {Z.~L.}\ \bibnamefont
  {Yuan}}, \bibinfo {author} {\bibfnamefont {B.}~\bibnamefont {Fröhlich}},
  \bibinfo {author} {\bibfnamefont {M.}~\bibnamefont {Lucamarini}}, \bibinfo
  {author} {\bibfnamefont {G.~L.}\ \bibnamefont {Roberts}}, \bibinfo {author}
  {\bibfnamefont {J.~F.}\ \bibnamefont {Dynes}}, \ and\ \bibinfo {author}
  {\bibfnamefont {A.~J.}\ \bibnamefont {Shields}},\ }\href {\doibase
  10.1103/PhysRevX.6.031044} {\bibfield  {journal} {\bibinfo  {journal} {Phys.
  Rev. X}\ }\textbf {\bibinfo {volume} {6}},\ \bibinfo {pages} {031044}
  (\bibinfo {year} {2016})}\BibitemShut {NoStop}%
\bibitem [{\citenamefont {Lin}\ \emph {et~al.}(2019)\citenamefont {Lin},
  \citenamefont {Upadhyaya},\ and\ \citenamefont {Lütkenhaus}}]{Lin:19}%
  \BibitemOpen
  \bibfield  {author} {\bibinfo {author} {\bibfnamefont {J.}~\bibnamefont
  {Lin}}, \bibinfo {author} {\bibfnamefont {T.}~\bibnamefont {Upadhyaya}}, \
  and\ \bibinfo {author} {\bibfnamefont {N.}~\bibnamefont {Lütkenhaus}},\
  }\href {\doibase 10.1103/PhysRevX.9.041064} {\bibfield  {journal} {\bibinfo
  {journal} {Phys. Rev. X}\ }\textbf {\bibinfo {volume} {9}},\ \bibinfo {pages}
  {041064} (\bibinfo {year} {2019})}\BibitemShut {NoStop}%
\bibitem [{\citenamefont {Laudenbach}\ \emph
  {et~al.}(2018{\natexlab{b}})\citenamefont {Laudenbach}, \citenamefont
  {Pacher}, \citenamefont {Fung}, \citenamefont {Poppe}, \citenamefont {Peev},
  \citenamefont {Schrenk}, \citenamefont {Hentschel}, \citenamefont {Walther},\
  and\ \citenamefont {H\"ubel}}]{Laudenbach:2018}%
  \BibitemOpen
  \bibfield  {author} {\bibinfo {author} {\bibfnamefont {F.}~\bibnamefont
  {Laudenbach}}, \bibinfo {author} {\bibfnamefont {C.}~\bibnamefont {Pacher}},
  \bibinfo {author} {\bibfnamefont {C.-H.~F.}\ \bibnamefont {Fung}}, \bibinfo
  {author} {\bibfnamefont {A.}~\bibnamefont {Poppe}}, \bibinfo {author}
  {\bibfnamefont {M.}~\bibnamefont {Peev}}, \bibinfo {author} {\bibfnamefont
  {B.}~\bibnamefont {Schrenk}}, \bibinfo {author} {\bibfnamefont
  {M.}~\bibnamefont {Hentschel}}, \bibinfo {author} {\bibfnamefont
  {P.}~\bibnamefont {Walther}}, \ and\ \bibinfo {author} {\bibfnamefont
  {H.}~\bibnamefont {H\"ubel}},\ }\href {\doibase 10.1007/s11432-022-3656-4}
  {\bibfield  {journal} {\bibinfo  {journal} {Adv. Quantum Technol.}\ }\textbf
  {\bibinfo {volume} {1}},\ \bibinfo {pages} {1800011} (\bibinfo {year}
  {2018}{\natexlab{b}})}\BibitemShut {NoStop}%
\bibitem [{\citenamefont {Lin}\ and\ \citenamefont
  {Lütkenhaus}(2020)}]{Lin:20}%
  \BibitemOpen
  \bibfield  {author} {\bibinfo {author} {\bibfnamefont {J.}~\bibnamefont
  {Lin}}\ and\ \bibinfo {author} {\bibfnamefont {N.}~\bibnamefont
  {Lütkenhaus}},\ }\href {\doibase 10.1103/PhysRevApplied.14.064030}
  {\bibfield  {journal} {\bibinfo  {journal} {Phys. Rev. Appl.}\ }\textbf
  {\bibinfo {volume} {14}},\ \bibinfo {pages} {064030} (\bibinfo {year}
  {2020})}\BibitemShut {NoStop}%
\bibitem [{\citenamefont {Winick}\ \emph {et~al.}(2018)\citenamefont {Winick},
  \citenamefont {Lütkenhaus},\ and\ \citenamefont {Coles}}]{Winick:18}%
  \BibitemOpen
  \bibfield  {author} {\bibinfo {author} {\bibfnamefont {A.}~\bibnamefont
  {Winick}}, \bibinfo {author} {\bibfnamefont {N.}~\bibnamefont {Lütkenhaus}},
  \ and\ \bibinfo {author} {\bibfnamefont {P.~J.}\ \bibnamefont {Coles}},\
  }\href {\doibase 10.2256/quantum.2.77} {\bibfield  {journal} {\bibinfo
  {journal} {Quantum}\ }\textbf {\bibinfo {volume} {2}},\ \bibinfo {pages} {77}
  (\bibinfo {year} {2018})}\BibitemShut {NoStop}%
\bibitem [{\citenamefont {Jeong}\ \emph {et~al.}(2022)\citenamefont {Jeong},
  \citenamefont {Jung},\ and\ \citenamefont {Ha}}]{Jeong:22}%
  \BibitemOpen
  \bibfield  {author} {\bibinfo {author} {\bibfnamefont {S.}~\bibnamefont
  {Jeong}}, \bibinfo {author} {\bibfnamefont {H.}~\bibnamefont {Jung}}, \ and\
  \bibinfo {author} {\bibfnamefont {J.}~\bibnamefont {Ha}},\ }\href {\doibase
  10.1038/s41534-021-00509-9} {\bibfield  {journal} {\bibinfo  {journal} {npj
  Quantum Inf.}\ }\textbf {\bibinfo {volume} {8}},\ \bibinfo {pages} {6}
  (\bibinfo {year} {2022})}\BibitemShut {NoStop}%
\bibitem [{\citenamefont {Feng}\ \emph {et~al.}(2023)\citenamefont {Feng},
  \citenamefont {Qiu}, \citenamefont {Zhang}, \citenamefont {Jiang},
  \citenamefont {Zhang}, \citenamefont {Huang},\ and\ \citenamefont
  {Zeng}}]{Feng:23}%
  \BibitemOpen
  \bibfield  {author} {\bibinfo {author} {\bibfnamefont {Y.}~\bibnamefont
  {Feng}}, \bibinfo {author} {\bibfnamefont {R.~H.}\ \bibnamefont {Qiu}},
  \bibinfo {author} {\bibfnamefont {K.}~\bibnamefont {Zhang}}, \bibinfo
  {author} {\bibfnamefont {X.~Q.}\ \bibnamefont {Jiang}}, \bibinfo {author}
  {\bibfnamefont {M.~X.}\ \bibnamefont {Zhang}}, \bibinfo {author}
  {\bibfnamefont {P.}~\bibnamefont {Huang}}, \ and\ \bibinfo {author}
  {\bibfnamefont {G.~H.}\ \bibnamefont {Zeng}},\ }\href {\doibase
  10.1007/s11432-022-3656-4} {\bibfield  {journal} {\bibinfo  {journal} {Sci.
  China Inf. Sci.}\ }\textbf {\bibinfo {volume} {66}},\ \bibinfo {pages}
  {180511} (\bibinfo {year} {2023})}\BibitemShut {NoStop}%
\bibitem [{\citenamefont {Yang}\ \emph {et~al.}(2017)\citenamefont {Yang},
  \citenamefont {Bai}, \citenamefont {Wang},\ and\ \citenamefont
  {Li}}]{Yang:17}%
  \BibitemOpen
  \bibfield  {author} {\bibinfo {author} {\bibfnamefont {S.~S.}\ \bibnamefont
  {Yang}}, \bibinfo {author} {\bibfnamefont {Z.~L.}\ \bibnamefont {Bai}},
  \bibinfo {author} {\bibfnamefont {X.~Y.}\ \bibnamefont {Wang}}, \ and\
  \bibinfo {author} {\bibfnamefont {Y.~M.}\ \bibnamefont {Li}},\ }\href
  {\doibase 10.1109/JPHOT.2017.2761807} {\bibfield  {journal} {\bibinfo
  {journal} {IEEE Photon. J.}\ }\textbf {\bibinfo {volume} {9}},\ \bibinfo
  {pages} {1} (\bibinfo {year} {2017})}\BibitemShut {NoStop}%
\bibitem [{\citenamefont {Chi}\ \emph {et~al.}(2011)\citenamefont {Chi},
  \citenamefont {Qi}, \citenamefont {Zhu}, \citenamefont {Qian}, \citenamefont
  {Lo}, \citenamefont {Youn}, \citenamefont {Lvovsky},\ and\ \citenamefont
  {Tian}}]{Chi:10}%
  \BibitemOpen
  \bibfield  {author} {\bibinfo {author} {\bibfnamefont {Y.~M.}\ \bibnamefont
  {Chi}}, \bibinfo {author} {\bibfnamefont {B.}~\bibnamefont {Qi}}, \bibinfo
  {author} {\bibfnamefont {W.}~\bibnamefont {Zhu}}, \bibinfo {author}
  {\bibfnamefont {L.}~\bibnamefont {Qian}}, \bibinfo {author} {\bibfnamefont
  {H.-K.}\ \bibnamefont {Lo}}, \bibinfo {author} {\bibfnamefont {S.-H.}\
  \bibnamefont {Youn}}, \bibinfo {author} {\bibfnamefont {A.~I.}\ \bibnamefont
  {Lvovsky}}, \ and\ \bibinfo {author} {\bibfnamefont {L.}~\bibnamefont
  {Tian}},\ }\href {\doibase 10.1007/s11432-022-3656-4} {\bibfield  {journal}
  {\bibinfo  {journal} {Quantum Inf. Comput.}\ }\textbf {\bibinfo {volume}
  {13}},\ \bibinfo {pages} {1} (\bibinfo {year} {2011})}\BibitemShut {NoStop}%
\bibitem [{\citenamefont {Kanitschar}\ \emph {et~al.}(2023)\citenamefont
  {Kanitschar}, \citenamefont {George}, \citenamefont {Lin}, \citenamefont
  {Upadhyaya},\ and\ \citenamefont {Lütkenhaus}}]{Kanitschar:23}%
  \BibitemOpen
  \bibfield  {author} {\bibinfo {author} {\bibfnamefont {F.}~\bibnamefont
  {Kanitschar}}, \bibinfo {author} {\bibfnamefont {I.}~\bibnamefont {George}},
  \bibinfo {author} {\bibfnamefont {J.}~\bibnamefont {Lin}}, \bibinfo {author}
  {\bibfnamefont {T.}~\bibnamefont {Upadhyaya}}, \ and\ \bibinfo {author}
  {\bibfnamefont {N.}~\bibnamefont {Lütkenhaus}},\ }\href {\doibase
  10.1103/PRXQuantum.4.040306} {\bibfield  {journal} {\bibinfo  {journal} {PRX
  Quantum}\ }\textbf {\bibinfo {volume} {4}},\ \bibinfo {pages} {040306}
  (\bibinfo {year} {2023})}\BibitemShut {NoStop}%
  \bibitem [{Sup()}]{Supplemental}%
  \BibitemOpen
  \href@noop {} {}\bibinfo {howpublished} {See Supplemental Material at
  \url{http://link.aps.org/supplemental/DOI} for results on heterodyne
  detections and data acquisition, phase drift, and performance comparion,
  which includes Refs. [9, 18, 46, 47, 49, 51, 54, 55]}\BibitemShut {NoStop}%
\end{thebibliography}%


\providecommand{\noopsort}[1]{}
\begin{thebibliography}{8}%
\makeatletter
\providecommand \@ifxundefined [1]{%
 \@ifx{#1\undefined}
}%
\providecommand \@ifnum [1]{%
 \ifnum #1\expandafter \@firstoftwo
 \else \expandafter \@secondoftwo
 \fi
}%
\providecommand \@ifx [1]{%
 \ifx #1\expandafter \@firstoftwo
 \else \expandafter \@secondoftwo
 \fi
}%
\providecommand \natexlab [1]{#1}%
\providecommand \enquote  [1]{``#1''}%
\providecommand \bibnamefont  [1]{#1}%
\providecommand \bibfnamefont [1]{#1}%
\providecommand \citenamefont [1]{#1}%
\providecommand \href@noop [0]{\@secondoftwo}%
\providecommand \href [0]{\begingroup \@sanitize@url \@href}%
\providecommand \@href[1]{\@@startlink{#1}\@@href}%
\providecommand \@@href[1]{\endgroup#1\@@endlink}%
\providecommand \@sanitize@url [0]{\catcode `\\12\catcode `\$12\catcode
  `\&12\catcode `\#12\catcode `\^12\catcode `\_12\catcode `\%12\relax}%
\providecommand \@@startlink[1]{}%
\providecommand \@@endlink[0]{}%
\providecommand \url  [0]{\begingroup\@sanitize@url \@url }%
\providecommand \@url [1]{\endgroup\@href {#1}{\urlprefix }}%
\providecommand \urlprefix  [0]{URL }%
\providecommand \Eprint [0]{\href }%
\providecommand \doibase [0]{http://dx.doi.org/}%
\providecommand \selectlanguage [0]{\@gobble}%
\providecommand \bibinfo  [0]{\@secondoftwo}%
\providecommand \bibfield  [0]{\@secondoftwo}%
\providecommand \translation [1]{[#1]}%
\providecommand \BibitemOpen [0]{}%
\providecommand \bibitemStop [0]{}%
\providecommand \bibitemNoStop [0]{.\EOS\space}%
\providecommand \EOS [0]{\spacefactor3000\relax}%
\providecommand \BibitemShut  [1]{\csname bibitem#1\endcsname}%
\let\auto@bib@innerbib\@empty
%</preamble>
\bibitem [{\citenamefont {Hajomer}\ \emph {et~al.}(2024)\citenamefont
  {Hajomer}, \citenamefont {Derkach}, \citenamefont {Jain}, \citenamefont
  {Chin}, \citenamefont {Andersen},\ and\ \citenamefont
  {Gehring}}]{Hajomer2024}%
  \BibitemOpen
  \bibfield  {author} {\bibinfo {author} {\bibfnamefont {A.~A.~E.}\
  \bibnamefont {Hajomer}}, \bibinfo {author} {\bibfnamefont {I.}~\bibnamefont
  {Derkach}}, \bibinfo {author} {\bibfnamefont {N.}~\bibnamefont {Jain}},
  \bibinfo {author} {\bibfnamefont {H.~M.}\ \bibnamefont {Chin}}, \bibinfo
  {author} {\bibfnamefont {U.~L.}\ \bibnamefont {Andersen}}, \ and\ \bibinfo
  {author} {\bibfnamefont {T.}~\bibnamefont {Gehring}},\ }\href {\doibase
  10.1126/sciadv.adi9474} {\bibfield  {journal} {\bibinfo  {journal} {Sci.
  Adv.}\ }\textbf {\bibinfo {volume} {10}},\ \bibinfo {pages} {eadi9474}
  (\bibinfo {year} {2024})}\BibitemShut {NoStop}%
\bibitem [{\citenamefont {Lu}\ \emph {et~al.}(2023)\citenamefont {Lu},
  \citenamefont {Wang}, \citenamefont {Zapatero}, \citenamefont {Chen},
  \citenamefont {Wang}, \citenamefont {Yin}, \citenamefont {Curty},
  \citenamefont {He}, \citenamefont {Wang}, \citenamefont {Chen},\ and\
  \citenamefont {et~al.}}]{Lu:23}%
  \BibitemOpen
  \bibfield  {author} {\bibinfo {author} {\bibfnamefont {F.~Y.}\ \bibnamefont
  {Lu}}, \bibinfo {author} {\bibfnamefont {Z.~H.}\ \bibnamefont {Wang}},
  \bibinfo {author} {\bibfnamefont {V.}~\bibnamefont {Zapatero}}, \bibinfo
  {author} {\bibfnamefont {J.~L.}\ \bibnamefont {Chen}}, \bibinfo {author}
  {\bibfnamefont {S.}~\bibnamefont {Wang}}, \bibinfo {author} {\bibfnamefont
  {Z.~Q.}\ \bibnamefont {Yin}}, \bibinfo {author} {\bibfnamefont
  {M.}~\bibnamefont {Curty}}, \bibinfo {author} {\bibfnamefont {D.~Y.}\
  \bibnamefont {He}}, \bibinfo {author} {\bibfnamefont {R.}~\bibnamefont
  {Wang}}, \bibinfo {author} {\bibfnamefont {W.}~\bibnamefont {Chen}}, \ and\
  \bibinfo {author} {\bibnamefont {et~al.}},\ }\href {\doibase
  10.1103/PhysRevLett.131.110802} {\bibfield  {journal} {\bibinfo  {journal}
  {Physical Review Letters}\ }\textbf {\bibinfo {volume} {131}},\ \bibinfo
  {pages} {110802} (\bibinfo {year} {2023})}\BibitemShut {NoStop}%
\bibitem [{\citenamefont {Hu}\ \emph {et~al.}(2023)\citenamefont {Hu},
  \citenamefont {Wang}, \citenamefont {Chan}, \citenamefont {Yuan},\ and\
  \citenamefont {Lo}}]{Hu:23}%
  \BibitemOpen
  \bibfield  {author} {\bibinfo {author} {\bibfnamefont {C.~Q.}\ \bibnamefont
  {Hu}}, \bibinfo {author} {\bibfnamefont {W.~Y.}\ \bibnamefont {Wang}},
  \bibinfo {author} {\bibfnamefont {K.-S.}\ \bibnamefont {Chan}}, \bibinfo
  {author} {\bibfnamefont {Z.~H.}\ \bibnamefont {Yuan}}, \ and\ \bibinfo
  {author} {\bibfnamefont {H.-K.}\ \bibnamefont {Lo}},\ }\href {\doibase
  10.1103/PhysRevLett.131.110801} {\bibfield  {journal} {\bibinfo  {journal}
  {Phys. Rev. Lett.}\ }\textbf {\bibinfo {volume} {131}},\ \bibinfo {pages}
  {110801} (\bibinfo {year} {2023})}\BibitemShut {NoStop}%
\bibitem [{\citenamefont {Ji}\ \emph {et~al.}(2024)\citenamefont {Ji},
  \citenamefont {Huang}, \citenamefont {Wang}, \citenamefont {Jiang},\ and\
  \citenamefont {Zeng}}]{Li:24}%
  \BibitemOpen
  \bibfield  {author} {\bibinfo {author} {\bibfnamefont {F.~Y.}\ \bibnamefont
  {Ji}}, \bibinfo {author} {\bibfnamefont {P.}~\bibnamefont {Huang}}, \bibinfo
  {author} {\bibfnamefont {T.}~\bibnamefont {Wang}}, \bibinfo {author}
  {\bibfnamefont {X.~Q.}\ \bibnamefont {Jiang}}, \ and\ \bibinfo {author}
  {\bibfnamefont {G.~H.}\ \bibnamefont {Zeng}},\ }\href {\doibase
  10.1364/PRJ.519909} {\bibfield  {journal} {\bibinfo  {journal} {Photonics
  Research}\ }\textbf {\bibinfo {volume} {12}},\ \bibinfo {pages} {1485}
  (\bibinfo {year} {2024})}\BibitemShut {NoStop}%
\bibitem [{\citenamefont {Qi}\ \emph {et~al.}(2020)\citenamefont {Qi},
  \citenamefont {Gunther}, \citenamefont {Evans}, \citenamefont {Williams},
  \citenamefont {Camacho},\ and\ \citenamefont {Peters}}]{Qi:20}%
  \BibitemOpen
  \bibfield  {author} {\bibinfo {author} {\bibfnamefont {B.}~\bibnamefont
  {Qi}}, \bibinfo {author} {\bibfnamefont {H.}~\bibnamefont {Gunther}},
  \bibinfo {author} {\bibfnamefont {P.~G.}\ \bibnamefont {Evans}}, \bibinfo
  {author} {\bibfnamefont {B.~P.}\ \bibnamefont {Williams}}, \bibinfo {author}
  {\bibfnamefont {R.~M.}\ \bibnamefont {Camacho}}, \ and\ \bibinfo {author}
  {\bibfnamefont {N.~A.}\ \bibnamefont {Peters}},\ }\href {\doibase
  10.1103/PhysRevApplied.13.054065} {\bibfield  {journal} {\bibinfo  {journal}
  {Physical Review Applied}\ }\textbf {\bibinfo {volume} {13}},\ \bibinfo
  {pages} {054065} (\bibinfo {year} {2020})}\BibitemShut {NoStop}%
\bibitem [{\citenamefont {Huang}\ \emph {et~al.}(2021)\citenamefont {Huang},
  \citenamefont {Wang}, \citenamefont {Chen}, \citenamefont {Wang},
  \citenamefont {Zhou},\ and\ \citenamefont {Zeng}}]{Huang:21}%
  \BibitemOpen
  \bibfield  {author} {\bibinfo {author} {\bibfnamefont {P.}~\bibnamefont
  {Huang}}, \bibinfo {author} {\bibfnamefont {T.}~\bibnamefont {Wang}},
  \bibinfo {author} {\bibfnamefont {R.}~\bibnamefont {Chen}}, \bibinfo {author}
  {\bibfnamefont {P.}~\bibnamefont {Wang}}, \bibinfo {author} {\bibfnamefont
  {Y.~M.}\ \bibnamefont {Zhou}}, \ and\ \bibinfo {author} {\bibfnamefont
  {G.~H.}\ \bibnamefont {Zeng}},\ }\href {\doibase 10.1088/1367-2630/ac3684}
  {\bibfield  {journal} {\bibinfo  {journal} {New Journal of Physics}\ }\textbf
  {\bibinfo {volume} {23}},\ \bibinfo {pages} {113028} (\bibinfo {year}
  {2021})}\BibitemShut {NoStop}%
\bibitem [{\citenamefont {Tian}\ \emph {et~al.}(2023)\citenamefont {Tian},
  \citenamefont {Zhang}, \citenamefont {Liu}, \citenamefont {Wang},
  \citenamefont {Lu}, \citenamefont {Wang},\ and\ \citenamefont
  {Li}}]{Tian:23}%
  \BibitemOpen
  \bibfield  {author} {\bibinfo {author} {\bibfnamefont {Y.}~\bibnamefont
  {Tian}}, \bibinfo {author} {\bibfnamefont {Y.}~\bibnamefont {Zhang}},
  \bibinfo {author} {\bibfnamefont {S.~S.}\ \bibnamefont {Liu}}, \bibinfo
  {author} {\bibfnamefont {P.}~\bibnamefont {Wang}}, \bibinfo {author}
  {\bibfnamefont {Z.~G.}\ \bibnamefont {Lu}}, \bibinfo {author} {\bibfnamefont
  {X.~Y.}\ \bibnamefont {Wang}}, \ and\ \bibinfo {author} {\bibfnamefont
  {Y.~M.}\ \bibnamefont {Li}},\ }\href {\doibase 10.1364/OL.492082} {\bibfield
  {journal} {\bibinfo  {journal} {Optics Letters}\ }\textbf {\bibinfo {volume}
  {48}},\ \bibinfo {pages} {2953} (\bibinfo {year} {2023})}\BibitemShut
  {NoStop}%
\bibitem [{\citenamefont {Wang}\ \emph {et~al.}(2022)\citenamefont {Wang},
  \citenamefont {Li}, \citenamefont {Pi}, \citenamefont {Pan}, \citenamefont
  {Shao}, \citenamefont {Ma}, \citenamefont {Zhang}, \citenamefont {Yang},
  \citenamefont {Zhang}, \citenamefont {Huang},\ and\ \citenamefont
  {et~al.}}]{Wang:22}%
  \BibitemOpen
  \bibfield  {author} {\bibinfo {author} {\bibfnamefont {H.}~\bibnamefont
  {Wang}}, \bibinfo {author} {\bibfnamefont {Y.}~\bibnamefont {Li}}, \bibinfo
  {author} {\bibfnamefont {Y.~D.}\ \bibnamefont {Pi}}, \bibinfo {author}
  {\bibfnamefont {Y.}~\bibnamefont {Pan}}, \bibinfo {author} {\bibfnamefont
  {Y.}~\bibnamefont {Shao}}, \bibinfo {author} {\bibfnamefont {L.}~\bibnamefont
  {Ma}}, \bibinfo {author} {\bibfnamefont {Y.~C.}\ \bibnamefont {Zhang}},
  \bibinfo {author} {\bibfnamefont {J.}~\bibnamefont {Yang}}, \bibinfo {author}
  {\bibfnamefont {T.}~\bibnamefont {Zhang}}, \bibinfo {author} {\bibfnamefont
  {W.}~\bibnamefont {Huang}}, \ and\ \bibinfo {author} {\bibnamefont
  {et~al.}},\ }\href {\doibase 10.1038/s42005-022-00941-z} {\bibfield
  {journal} {\bibinfo  {journal} {Communications Physics}\ }\textbf {\bibinfo
  {volume} {5}},\ \bibinfo {pages} {162} (\bibinfo {year} {2022})}\BibitemShut
  {NoStop}%
\end{thebibliography}%
\end{document}

\preprint{APS/123-QED}
	
	\title{Supplemental Materials: High-Performance Fully Passive Discrete-State Continuous-Variable Quantum Key Distribution With Local Local Oscillator}
	
	\author{Yu Zhang}
	\affiliation{State Key Laboratory of Quantum Optics Technologies and Devices, Institute of Opto-Electronics, Shanxi University, Taiyuan 030006, China}
	\author{Xuyang Wang}
	\email{wangxuyang@sxu.edu.cn}
	\affiliation{State Key Laboratory of Quantum Optics Technologies and Devices, Institute of Opto-Electronics, Shanxi University, Taiyuan 030006, China}
	\affiliation{Collaborative Innovation Center of Extreme Optics, Shanxi University, Taiyuan 030006, China}
	\affiliation{Hefei National Laboratory, Hefei 230088, China}
	\author{Chenyang Li}
	\email{chenyangli@astri.org}
	\affiliation{Hong Kong Applied Science and Technology Research Institute}
	\author{Jie Yun}
	\affiliation{State Key Laboratory of Quantum Optics Technologies and Devices, Institute of Opto-Electronics, Shanxi University, Taiyuan 030006, China}
	\author{Qiang Zeng}
	\affiliation{Beijing Academy of Quantum Information Sciences, Beijing 100193, China}
	\author{Zhiliang Yuan}
	\affiliation{Beijing Academy of Quantum Information Sciences, Beijing 100193, China}
	\author{Zhenguo Lu}
	\affiliation{State Key Laboratory of Quantum Optics Technologies and Devices, Institute of Opto-Electronics, Shanxi University, Taiyuan 030006, China}
	\affiliation{Collaborative Innovation Center of Extreme Optics, Shanxi University, Taiyuan 030006, China}
	\author{Yongmin Li}
	\email{yongmin@sxu.edu.cn}
	\affiliation{State Key Laboratory of Quantum Optics Technologies and Devices, Institute of Opto-Electronics, Shanxi University, Taiyuan 030006, China}
	\affiliation{Collaborative Innovation Center of Extreme Optics, Shanxi University, Taiyuan 030006, China}
	\affiliation{Hefei National Laboratory, Hefei 230088, China}

	\date{\today}
	
	\maketitle
	
	\section{I. The heterodyne detection and data acquisition}
	
	Three heterodyne detections ($\text{HDs}$) were employed in our experiment: $\text{HD}_1$ with two pulsed beams, $\text{HD}_2$ with one strong pulsed beam and one continuous local local oscillator (LLO) beam, and $\text{HD}_3$ with one pulsed quantum signal beam and one continuous LLO beam. The bandwidths of homodyne detectors in $\text{HD}_1$ and $\text{HD}_2$ were 5 GHz and that in $\text{HD}_3$ was 23 GHz. A high-speed digital oscilloscope was employed to acquire the quadrature signals of $\text{HD}_3$ with 5 GHz bandwidth and 50 Gsample/s sampling rate. 
	\begin{figure}[htbp]
		\centering
		\includegraphics[scale=0.15]{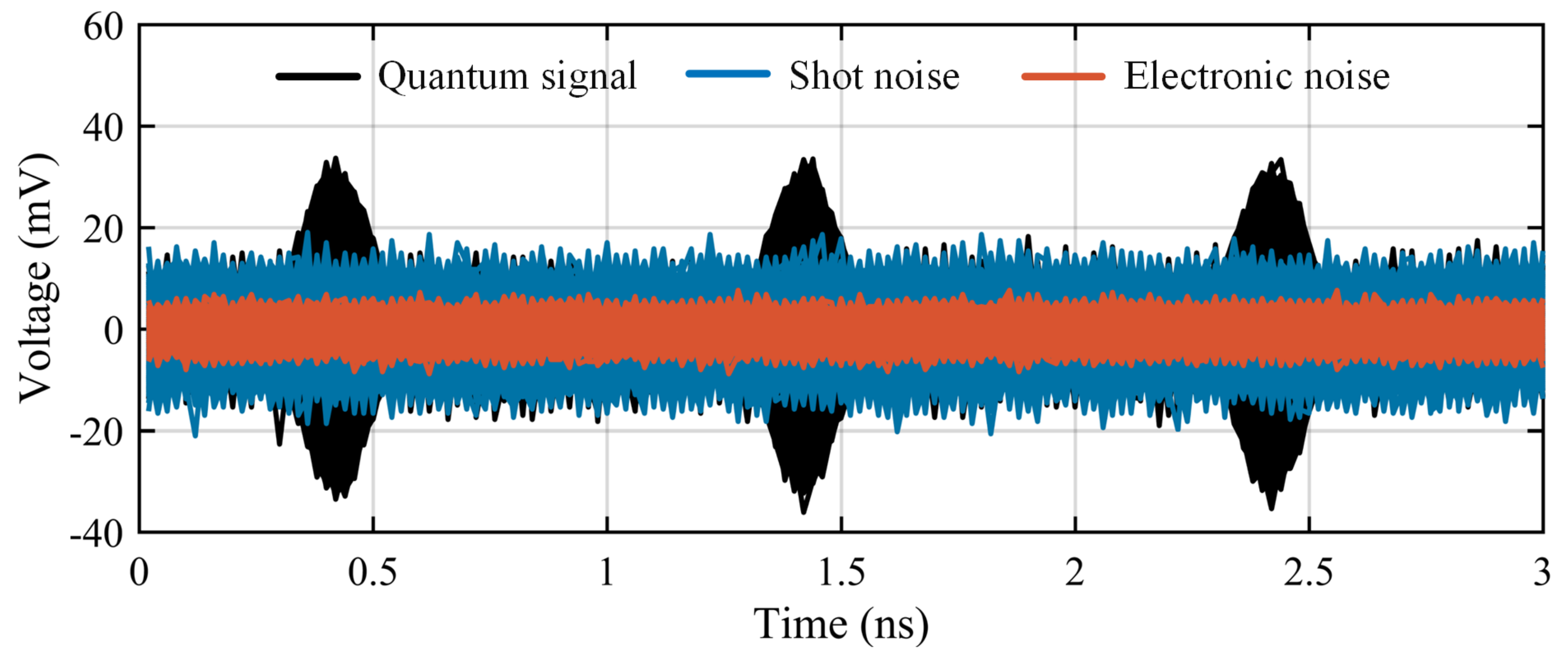}% Here is how to import EPS art
		\caption{The time traces of various outputs of $\text{HD}_3$.}
		\label{fig:1}
	\end{figure}
	
	Fig.~\ref{fig:1} presents the time traces of various outputs of $\text{HD}_3$, namely, the electronic noise (orange traces), shot noise (blue traces), and the quadrature signals (black traces) with $V_{\mathrm{A}} = 10$. The variances of the shot noise and electronic noise is $\sigma^2 = 1.58 \times 10^{-5} \text{ V}^2$, and $V_{el} = 2.99 \times 10^{-6} \ \, \mathrm{V}^2$ ($0.189~\text{SNU}$), respectively.
	
	Given that the detection efficiency of the homodyne detector in $\text{HD}_3$ is $\eta_{d} = 0.436$, the transmission efficiency of the 90$^\circ$ optical hybrid is $\eta_{h} = 0.853$, the transmission efficiency of the dynamic polarization controller and PBS is $\eta_{p} = 0.946$, and the total transmission efficiency at the receiver side is $\eta = \eta_d \cdot \eta_h \cdot \eta_p = 0.352$.
	\begin{figure}[htbp]
		\centering
		\includegraphics{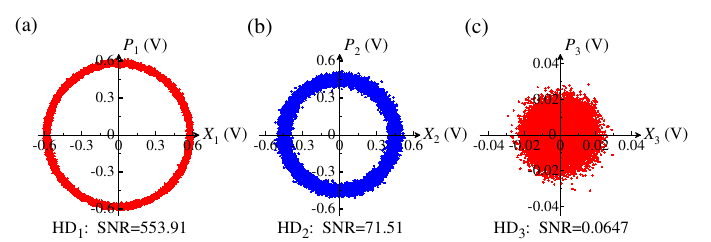}
		\caption{Distribution of raw measurement data of the three $\text{HDs}$ in phase space.}
		\label{fig:2}
	\end{figure}
	
	The raw measurement data of the three $\text{HDs}$ are presented in Fig.~\ref{fig:2} when the transmission length is 25 km. The SNR of $\text{HD}_1$, $\text{HD}_2$, and $\text{HD}_3$ are 553.91, 71.51, and 0.0647, by using the definition
	\begin{equation}
		\text{SNR} = \frac{\bar{R}^2}{V_R},
		\label{eq:S1}
	\end{equation}
	where $\bar{R} = {\sum_{i=1}^{N} {R_i}} /N $, $R_i = \sqrt{X_i^2 + P_i^2}$, denotes the average radius of circle in Fig.~\ref{fig:2}(a) and (b), and $V_R = {\sum_{i=1}^{N} {(R_i - \bar{R})^2}} /N $ denotes the variance of the radius.
	
	The finite SNR of $\text{HD}_1$ and $\text{HD}_2$ results in the estimate errors of the relative phases and induces the excess noise to the CV-QKD system. On Alice’s side, the induced excess noise $\varepsilon_{1}$  due to the finite $\text{SNR}_{1}$ of $\text{HD}_1$ is given by
	\begin{equation}
		{{\varepsilon }_{1}}=\sigma _{{{\Phi }_{1}}}^{2}\cdot 2{{\left| \alpha  \right|}^{2}},\quad {{\sigma }_{{{\Phi }_{1}}}}=1/\sqrt{2\cdot \text{SNR}_{1}}, 
		\label{eq:S2}
	\end{equation}
	where ${{\sigma }_{{{\Phi }_{1}}}}$ is the standard deviation of the phase estimation for ${{\Phi }_{1}}$, $\left| \alpha  \right|$ is the amplitude of the prepared quantum signals.
	
	Similarly, on Bob’s side, the induced excess noise ${{\varepsilon }_{2}}$ due to the finite $\text{SNR}_{2}$ of $\text{HD}_2$ is given by
	\begin{equation}
	{{\varepsilon }_{2}}=\sigma _{{{\Phi }_{2}}}^{2}\cdot 2{{\left| \alpha  \right|}^{2}},\quad  {{\sigma }_{{{\Phi }_{2}}}}=1/\sqrt{2\cdot \text{SNR}_{2}},
	\label{eq:S3}
	\end{equation}
	where ${{\sigma }_{{{\Phi }_{2}}}}$ is the standard deviation of the phase estimation for ${{\Phi }_{2}}$. To derive the induced excess noise in Eq. (\ref{eq:S2}), a linear quantum channel is assumed. A higher intensity of reference pulses means a higher $\text{SNR}_{2}$, however, strong reference pulses will increase the excess noise due to the leakage and nonlinear scattering. Therefore, a reasonable SNR of $\text{HD}_2$ are selected at each transmission length. 

	For estimation of the slow phase drift of each data block, there are still minor residual errors that can be approximately given by
	\begin{equation}
	\sigma _{{{\theta }_{\max }}}=\sqrt{\frac{\sum\limits_{i=1}^{N-1}{{{\left( {{\theta }_{\max }}\left( i+1 \right)-{{\theta }_{\max }}\left( i \right) \right)}^{2}}}}{N-1}}, 
	\label{eq:S4}
	\end{equation}
	where ${{\theta }_{\max }}\left( i \right)$ represents the compensation phase of the \textit{i}th block. This residual phase error $\sigma _{{{\theta }_{\max }}}$ can cause excess noise
	\begin{equation}
		{{\varepsilon }_{3}}={\sigma _{{{\theta }_{\max }}}}^{2}\cdot 2{{\left| \alpha  \right|}^{2}}.
		\label{eq:S5}
	\end{equation}
	The residual phase error $\sigma _{{{\theta }_{\max }}}$ mainly arises from the precision of estimation and the duration time of each data block. If more data are used to estimating ${{\theta }_{\max }}$, the statistical fluctuation effect can be suppressed and a higher precision will be achieved. However, the duration time of phase drift will be longer, which increase the phase fluctuation. In our experiment, a block length of $10^{5}$ is selected to get the best balance.

	By inserting the measured $\text{SNR}_{1}$, $\text{SNR}_{2}$, and $\sigma _{{{\theta }_{\max }}}$into Eqs. (\ref{eq:S2}), (\ref{eq:S3}), and (\ref{eq:S4}), the induced excess noises are calculated to be ${{\varepsilon }_{1}}=1.22\times10^{-3}~\text{SNU}$, ${{\varepsilon }_{2}}=9.43\times10^{-3}~\text{SNU}$, and ${{\varepsilon }_{3}}=1.67\times10^{-5}~\text{SNU}$. We find that the finite $\text{SNR}_{2}$ is a key factor that contribute to the excess noise.
	\begin{figure}[htbp]
		\centering
		\includegraphics[width=\columnwidth]{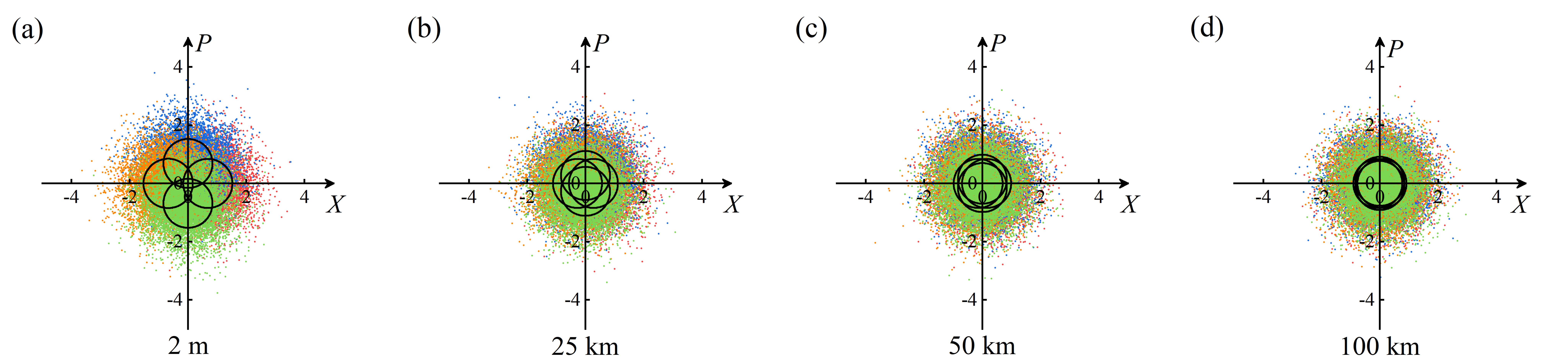}
		\caption{Displaced thermal states in phase space at different transmission lengths: (a) 2 m; (b) 25 km; (c) 50 km; and (d) 100 km.}
		\label{fig:3}
	\end{figure}

	Here, we focus on the processing of quadratures of $\text{HD}_3$. After phase rotation and diseretization, the quadratures of the four displaced thermal states $\rho_{\text{th}k},~k \in \{0,1,2,3\}$ in phase space at different transmission lengths are presented in Fig.~\ref{fig:3}. The red, blue, orange, and green dots represent the quadratures of the states $\rho_{\text{th}0}$, $\rho_{\text{th}1}$, $\rho_{\text{th}2}$, and $\rho_{\text{th}3}$, respectively, and the black circles denote the error circles. The average photon number of the displaced thermal states at the transmission lengths of $2$ m, $25$ km, $50$ km, and $100$ km were $0.3275$, $0.1325$, $0.1050$, and $0.0925$, respectively.

	During the experiment, the secret key rates of 50 groups of pulses were calculated at each transmission length. The first- and second-order moments are displayed in Figs.~\ref{fig:4}--\ref{fig:7}, and the subgraphs (a), (b), (c), and (d) represent the states $\rho_{\text{th}0}$, $\rho_{\text{th}1}$, $\rho_{\text{th}2}$, and $\rho_{\text{th}3}$, respectively. In each figure, the hollow triangles and solid triangles represent the first-order moment $\langle \hat{X}_{k} \rangle$ and second-order moment $\langle \hat{X}_{k}^{2} \rangle$ of the quadratures $\hat{X}_{k}$, while the hollow squares and solid squares represent the first-order moment $\langle \hat{P}_{k} \rangle$ and second-order moment $\langle \hat{P}_{k}^{2} \rangle$ of the quadratures $\hat{P}_{k}$. The first- and second-order moments have been normalized to the standard deviation $\sigma$ and the variance $\sigma^{2}$ of the shot noise, respectively.
	\begin{figure*}[htbp]
		\centering
		\includegraphics[scale=0.06]{FigS4.pdf}
		\caption{First- and second-order moments of the quadratures of the displaced thermal states at $2$ m.}
		\label{fig:4}
	\end{figure*}
	\begin{figure*}[htbp]
		\centering
		\includegraphics[scale=0.06]{FigS5.pdf}
		\caption{First- and second-order moments of the quadratures of the displaced thermal states at $25$ km.}
		\label{fig:5}
	\end{figure*}
	\begin{figure*}[htbp]
		\centering
		\includegraphics[scale=0.06]{FigS6.pdf}
		\caption{First- and second-order moments of the quadratures of the displaced thermal states at $50$ km.}
		\label{fig:6}
	\end{figure*}
	\begin{figure*}[htbp]
		\centering
		\includegraphics[scale=0.06]{FigS7.pdf}
		\caption{First- and second-order moments of the quadratures of the displaced thermal states at $100$ km.}
		\label{fig:7}
	\end{figure*}

	\section{II. Phase drift}
	
\begin{figure}[htbp]
		\centering
		\includegraphics[scale=0.13]{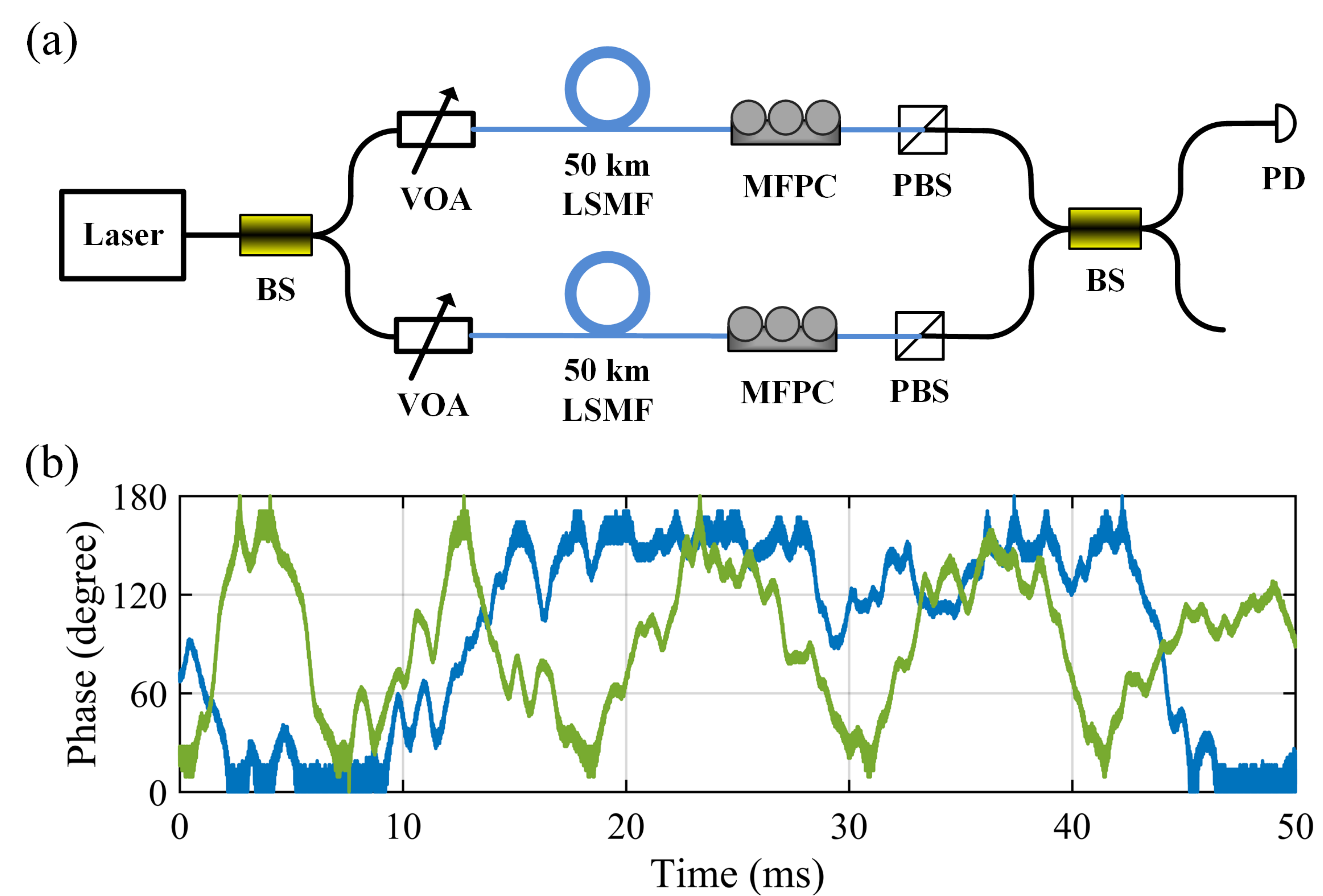}% Here is how to import EPS art
		\caption{Experimental setup (a) and results (b) for the phase drift arising from the 100 km LSMF. BS: beam splitter; VOA: variable optical attenuator; 50 km LSMF: 50 kilometer low-loss single-mode fiber; MFPC: manual fiber polarization controller; PBS: polarization beam splitter; PD: photodetector.}
		\label{fig:8}
	\end{figure}

	The phase drift due to the 100 km low-loss single-mode fiber (LSMF) was characterized using the experimental setup shown in Fig.~\ref{fig:8}(a). A laser with a narrow linewidth of $100$ Hz and wavelength of $1550.12$ nm was employed to minimize the effect of phase noise of laser. The laser was split into two portions, and each passed through a $50$ km LSMF. Prior to interference on a $50/50$ beam splitter, their polarization was corrected using polarization controllers and polarization beam splitters. One of the output beams was detected by a photodetector with bandwidth of $200$ kHz.
	
	As shown in Fig.~\ref{fig:8}(b), the phase drift due to the 100 km fiber is random, which is mainly caused by temperature variations or vibrations in external environment. Two typical results are plotted, a slow drift (blue line) and a fast drift (green line). From the measurement outcomes, we can see that it takes at least 2.1 ms to transition from the phase $\pi$ to the phase $0$. This corresponds to a phase drift speed of about 1.5 krad/s. In this case, the phase drift in each quantum signal pulse (with duration time of 200 ps) is extremely small and can be neglected.

\begin{figure}[htbp]
		\centering
		\includegraphics[scale=0.03]{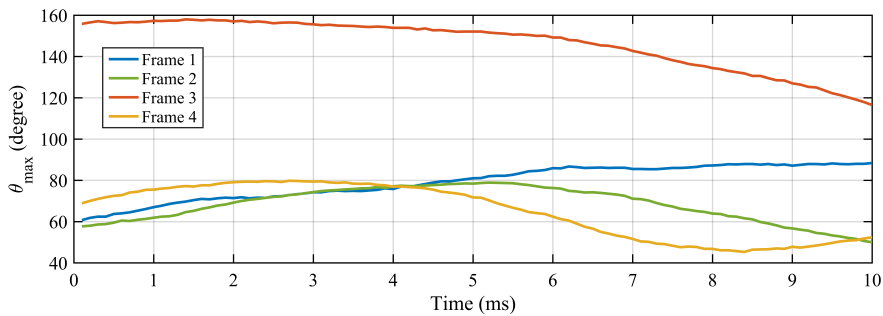}
		\caption{The compensation phases $\theta_{\text{max}}$ of blocks in four frames.}
		\label{fig:9}
	\end{figure}

	Figure~\ref{fig:9} presents the compensation phases $\theta_{\text{max}}$ of blocks in four frames. There are a total of 100 blocks in each frame. The maximum phase drifts between two adjacent blocks in frames 1, 2, 3, and 4 are $0.6^\circ$, $0.9^\circ$, $0.9^\circ$, and $1.1^\circ$, respectively. It corresponds to a maximum phase drift speed of 175 rad/s, which is mainly caused by the short fibers (tens of meters) in Alice's and Bob's systems.

	\section{III. Performance comparison}
	
	Table~\ref{tab:1} lists the comparison of the performance of our system to those of previous high-performance passive QKD protocols and modulated CV-QKD protocols. Our system achieves the maximum transmission length as that of the Gaussian modulation (GM) LLO CV-QKD protocol, whereas our secret key bit rate is five times higher than its~\cite{Hajomer2024}. The repetition rate and secret key rate are one order higher than those of the passive discrete-variable (DV) QKD protocols, furthermore, the transmission length is longer than theirs~\cite{Lu:23,Hu:23}. The secret key rate is also comparable to the discreted-modulation (DM) and passive thermal state CV-QKD protocols, but with a longer transmission length~\cite{Li:24,Qi:20,Huang:21,Tian:23,Wang:22}. 
	\begin{table*}[htbp]
		\caption{Performance comparison between our protocol and previous passive and modulated protocols. TLO: transmitted local oscillator; LCA: linear channel assumption; SDP: semidefinite programming.}
		\label{tab:1}
		\small
		\begin{ruledtabular}	
			\begin{tabular}{cccccccc}
				${\text{Literature}}$ & ${\text{Protocol}}$ & \multicolumn{2}{c}{Repetition rate} & ${L~\text{(km)}}$ & ${Channel~loss~\text{(dB)}}$ & ${R~\text{(bits/pulse)}}$ & ${K~\text{(Mbps)}}$ \\ 
				\hline
				\multirow{2}{*}{\cite{Lu:23}} & \multirow{2}{*}{${\text{Passive DV}}$} & \multicolumn{2}{c}{\multirow{2}{*}{200 MHz}} & $\text{30}$ & -- & $3.3\times10^{-4}$ &-- \\ 
				\cline{5-8}
				& & \multicolumn{2}{c}{} & $\text{50}$ & -- & $9.9\times10^{-5}$ & -- \\ 
				\hline
				\multirow{3}{*}{\cite{Hu:23}} & \multirow{3}{*}{${\text{Passive DV}}$} & \multicolumn{2}{c}{\multirow{3}{*}{20 MHz}} & -- & $7.2$ & $7.65\times10^{-5}$ & --\\ \cline{5-8}
				& & \multicolumn{2}{c}{} & -- & $11.6$ & $4.01\times10^{-5}$ & -- \\ \cline{5-8}
				& & \multicolumn{2}{c}{} & -- & $16.7$ & $1.86\times10^{-5}$ & -- \\ \cline{5-8}
				\hline
				\cite{Li:24} & ${\text{Passive thermal state CV TLO}}$ & \multicolumn{2}{c}{23 GHz} & ${\text{5}}$ & -- & -- & ${\text{1090}}$ \\ 
				\hline	
				\multirow{2}{*}{\cite{Qi:20}} & \multirow{2}{*}{${\text{Passive thermal state CV TLO}}$} & \multicolumn{2}{c}{\multirow{2}{*}{--}} & $\text{8.05}$ & -- & $\sim 10^{-2}$ & -- \\ 		
				\cline{5-8}
				& & \multicolumn{2}{c}{} & $\text{41.2}$ & -- & $\sim 10^{-3}$ & -- \\ 
				\hline
				\cite{Huang:21} & ${\text{Passive thermal state CV TLO}}$ & \multicolumn{2}{c}{350 MHz} & ${\text{5}}$ & -- & -- & ${\text{9.3}}$ \\ 
				\hline
				\multirow{4}{*}{${\text{Current work}}$} & \multirow{4}{*}{${\text{Passive discrete-state CV LLO}}$} & \multicolumn{2}{c}{\multirow{4}{*}{1 GHz}} & $\text{0.002}$& --  & $1.11\times10^{-1}$ & $\text{89.9}$ \\ 	
				\cline{5-8}
				& & \multicolumn{2}{c}{} & $\text{25}$ & -- & $6.73\times10^{-3}$ & $\text{5.45}$ \\ \cline{5-8}
				& & \multicolumn{2}{c}{} & $\text{50}$ & -- & $9.33\times10^{-4}$ & $\text{0.756}$ \\ \cline{5-8}
				& & \multicolumn{2}{c}{} & $\text{100}$ & -- & $1.57\times10^{-4}$ & $\text{0.127}$\\ 
				\hline
				\cite{Hajomer2024} & ${\text{GM CV LLO}}$ & \multicolumn{2}{c}{100 Mbaud} & $\text{100}$ & -- & -- & $\text{$\geq 0.025$}$ \\ 
				\hline
				\multirow{3}{*}{\cite{Tian:23}} & \multirow{3}{*}{${\text{DM CV LLO}}$} & \multicolumn{2}{c}{\multirow{3}{*}{2.5 Gbaud}} & $\text{25}$ & -- & -- & $\text{49.02}$ \\ 		
				\cline{5-8}
				& & \multicolumn{2}{c}{} & $\text{50}$ & -- & -- & $\text{11.86}$ \\ \cline{5-8}
				& & \multicolumn{2}{c}{} & $\text{80}$ & -- & -- & $\text{2.11}$  \\ 
				\hline
				\multirow{6}{*}{\cite{Wang:22}} & \multirow{6}{*}{${\text{DM CV LLO}}$} & \multirow{6}{*}{${\text{5 Gbaud}}$} & \multirow{3}{*}{${\text{LAC}}$} &  $\text{5}$ & -- & -- & $\text{190.54}$ \\ 	
				\cline{5-8}
				& & & & $\text{10}$ & -- & -- & $\text{137.36}$ \\ \cline{5-8}
				& & & & $\text{25}$ & -- & -- & $\text{52.48}$  \\ \cline{4-8}
				& & & \multirow{3}{*}{${\text{SDP}}$} & $\text{5}$ & -- & -- & $\text{233.87}$ \\  
				\cline{5-8}
				& & & & $\text{10}$ & -- & -- & $\text{133.6}$ \\ \cline{5-8}
				& & & & $\text{25}$ & -- & -- & $\text{21.53}$  \\ 	
			\end{tabular}
		\end{ruledtabular}
	\end{table*}
	\bibliography{SupplementalMaterial}